\documentclass[pre,aps,10pt,superscriptaddress,twocolumn,floatfix]{revtex4-1}

\usepackage[nice]{nicefrac}
\usepackage{graphicx,color}
\usepackage{amsmath,amssymb,bm,bbm}
\usepackage{amsmath}
\usepackage{braket}
\usepackage{float}

\usepackage[plainpages=false,pdfpagelabels,colorlinks=true,linkcolor=red,urlcolor=blue,citecolor=blue,pdftitle={},pdfauthor={},pdfdisplaydoctitle=true,pdfduplex=DuplexFlipLongEdge]{hyperref}

\definecolor{darkred}{rgb}{0.90,0.2,0.2}
\definecolor{darkgreen}{rgb}{0,0.60,.2}
\definecolor{darkblue}{rgb}{0.1,0.3,1}
\definecolor{grey}{cmyk}{0,0,0,0.25}
\definecolor{orange}{cmyk}{0,0.6,0.8,0}

\begin{document}

\title{Single-particle eigenstate thermalization in quantum-chaotic quadratic Hamiltonians}

\author{Patrycja  \L yd\.{z}ba}
\affiliation{Department of Theoretical Physics, J. Stefan Institute, SI-1000 Ljubljana, Slovenia}
\affiliation{Department of Theoretical Physics, Wroclaw University of Science and Technology, 50-370 Wroc{\l}aw, Poland}
\author{Yicheng Zhang}
\affiliation{Department of Physics, The Pennsylvania State University, University Park, Pennsylvania 16802, USA}
\author{Marcos Rigol}
\affiliation{Department of Physics, The Pennsylvania State University, University Park, Pennsylvania 16802, USA}
\author{Lev Vidmar}
\affiliation{Department of Theoretical Physics, J. Stefan Institute, SI-1000 Ljubljana, Slovenia}
\affiliation{Department of Physics, Faculty of Mathematics and Physics, University of Ljubljana, SI-1000 Ljubljana, Slovenia}

\begin{abstract}
We study the matrix elements of local and nonlocal operators in the single-particle eigenstates of two paradigmatic quantum-chaotic quadratic Hamiltonians; the quadratic Sachdev-Ye-Kitaev (SYK2) model and the three-dimensional Anderson model below the localization transition. We show that they display eigenstate thermalization for normalized observables. Specifically, we show that the diagonal matrix elements exhibit vanishing eigenstate-to-eigenstate fluctuations and that their variance is proportional to the inverse Hilbert space dimension. We also demonstrate that the ratio between the variance of the diagonal and the off-diagonal matrix elements is $2$, as predicted by the random matrix theory. We study distributions of matrix elements of observables and establish that they need not be Gaussian. We identify the class of observables for which the distributions are Gaussian.
\end{abstract}
\maketitle

\section{Introduction}

Whether isolated quantum many-body systems thermalize after being taken far from equilibrium has fascinated researchers since the early days of quantum mechanics~\cite{vonneumann_29}. It has been experimentally demonstrated in several ultracold-gas quantum simulators that they do under certain conditions~\cite{Trotzky2012, Kaufman2016, clos_porras_16, tang_kao_18}. On the theory side, we have learned that thermalization occurs generically in many-body interacting (quantum-chaotic) systems, and that quantum chaos can be identified, among other ways, by the following properties of the many-body eigenenergies and eigenstates of the Hamiltonian:
(i) the statistics of energy levels agrees with the predictions of the random matrix theory (RMT)~\cite{bohigas_giannoni_84, montambaux_poilblanc_93, hsu_dauriac_93, poilblanc_ziman_93, distasio_zotos_95, prosen_99, santos_04, rabson_narozhny_04, kolovsky_buchleitner_04, santos_rigol_10a, santos_rigol_10b, kollath_roux_10},
(ii) the matrix elements of observables in energy eigenstates comply with the eigenstate thermalization hypothesis (ETH)~\cite{deutsch_91, srednicki_94, srednicki_99, rigol_dunjko_08, dalessio_kafri_16, mori_ikeda_18, deutsch_18},
and (iii) the structure of energy eigenstates is chaotic~\cite{santos_rigol_10a} as manifested by, e.g., a maximal volume-law entanglement entropy~\cite{deutsch_10, santos_polkovnikov_12, deutsch_li_2013, vidmar_rigol_17, garrison_grover_18, dymarsky_lashkari_18, huang_19, murthy_srednicki_19a, miao_barthel_21}.

While the matrix elements of observables have been widely studied computationally in lattice models in many-body eigenstates of quantum-chaotic Hamiltonians~\cite{rigol_dunjko_08, rigol_09a, rigol09, santos_rigol_10b, steinigeweg_herbrych_13, khatami_pupillo_13, beugeling_moessner_14, sorg14, steinigeweg_khodja_14, kim_ikeda_14, beugeling_moessner_15, mondaini_fratus_16, luitz_16, mondaini_rigol_17, yoshizawa_iyoda_18, khaymovich_haque_19, jansen_stolpp_19, leblond_mallayya_19, mierzejewski_vidmar_20, brenes_leblond_20, brenes_goold_20, noh_sagawa_20, richter_dymarsky_20, leblond_rigol_20, sugimoto_hamazaki_21, noh_21, schoenle_jansen_21, fritsch_prosen_21}, we are not aware of parallel studies in single-particle eigenstates of quantum-chaotic quadratic Hamiltonians. We stress that we refer to Hamiltonians of interacting systems for which the many-body spectrum exhibits quantum chaos as quantum-chaotic interacting Hamiltonians, and to quadratic Hamiltonians for which the single-particle spectrum exhibits quantum chaos as quantum-chaotic quadratic Hamiltonians~\cite{lydzba_rigol_21}. Examples of quantum-chaotic quadratic models in a lattice include the three-dimensional Anderson model below the localization transition~\cite{altshuler_shklovskii_86, altshuler_zharekeshev_88, shklovskii_shapiro_93, hofstetter_schreiber_93, sierant_delande_20, suntajs_prosen_21} and the quadratic SYK2 model~\cite{lydzba_rigol_20, lydzba_rigol_21, liu_chen_18}. For the latter, the agreement with the RMT predictions is guaranteed by construction. Our goal in this work is to explore the properties of matrix elements of observables in {\it single-particle} eigenstates of quantum-chaotic quadratic Hamiltonians, as well as to identify similarities and differences with the properties of matrix elements of observables in many-body eigenstates of quantum-chaotic interacting systems.

We focus on the previously mentioned examples of quantum-chaotic quadratic Hamiltonians; the quadratic SYK2 model in its Dirac fermion formulation and the three-dimensional (3D) Anderson model below the localization transition. We study the matrix elements of observables in the single-particle energy eigenstates $\{|\alpha\rangle\}$, where $\hat H |\alpha\rangle = E_{\alpha} |\alpha\rangle$ and $E_\alpha$ is the eigenenergy corresponding to $|\alpha\rangle$. We show that properly normalized observables [with a unit Hilbert-Schmidt norm, see Eq.~(\ref{def_norm})] exhibit eigenstate thermalization. Specifically, we show that:
(i) For the diagonal matrix elements, the average eigenstate-to-eigenstate fluctuations decrease $\propto 1/\sqrt{V}$ while the variance decreases $\propto 1/V$, where $V$ is the number of lattice sites and hence the dimension of the single-particle Hilbert space. Similar scalings are observed in quantum-chaotic interacting systems after replacing $V\rightarrow{\cal D}$, where $\cal D$ is the dimension of the many-body Hilbert space~\cite{beugeling_moessner_14, kim_ikeda_14, mondaini_fratus_16, mondaini_rigol_17, yoshizawa_iyoda_18, jansen_stolpp_19, leblond_mallayya_19, mierzejewski_vidmar_20, sugimoto_hamazaki_21, richter_dymarsky_20, leblond_rigol_20, noh_21, haque_mcclarty_19}.
(ii) The ratio between the variance of diagonal and off-diagonal matrix elements is $2$, as predicted by the RMT~\cite{dalessio_kafri_16}. Such a ratio has been observed in quantum-chaotic interacting systems~\cite{mondaini_rigol_17, jansen_stolpp_19, richter_dymarsky_20, schoenle_jansen_21}.

For the matrix elements of an observable $\hat O$ in the single-particle eigenstates of quantum-chaotic quadratic Hamiltonians, the ETH ansatz~\cite{srednicki_99, dalessio_kafri_16} can be written as
\begin{equation} \label{def_eth_ansatz}
\langle \alpha |\hat O |\beta\rangle = {\cal O}(\bar E) \delta_{\alpha\beta} + \rho(\bar E)^{-1/2} {\cal F}(\bar E, \omega) R_{\alpha\beta}\;,
\end{equation}
where $\bar E = (E_\alpha + E_\beta)/2$, $\omega = E_\beta-E_\alpha$, ${\cal O}(\bar E)$ and ${\cal F}(\bar E, \omega)$ are smooth functions of their arguments, and $\rho(\bar E) = \delta N/\delta E|_{\bar E}$ is the single-particle density of states at energy $\bar E$. The latter typically scales as $V$. The distribution of matrix elements is described by the random variable $R_{\alpha\beta}$, which has zero mean and unit variance.
For observables studied in quantum-chaotic interacting models on a lattice, the distribution of matrix elements has been found to be Gaussian~\cite{beugeling_moessner_15, luitz_barlev_16, khaymovich_haque_19, leblond_mallayya_19, brenes_leblond_20, brenes_goold_20, leblond_rigol_20, santos_perezbernal_20, noh_21, brenes_pappalardi_21}. Here we show that the distribution of matrix elements for observables in single-particle eigenstates of quantum-chaotic quadratic models need not be Gaussian. One of our goals is to identify which classes of single-particle observables exhibit Gaussian versus non-Gaussian distributions, and to understand the origin of the difference with their many-body counterparts in quantum-chaotic interacting Hamiltonians.

The presentation is organized as follows. In Sec.~\ref{sec:models}, we introduce the models and observables under investigation. We define two ``versions'' of each observable: (i) the traditionally known version, which is measured in experiments involving many-particle systems, and (ii) the version that has a unit Hilbert-Schmidt norm in the single-particle Hilbert space (the normalized version). In Sec.~\ref{sec:structure}, we study the behavior of diagonal and off-diagonal matrix elements of these observables in the single-particle eigenstates of the Hamiltonians of interest. We focus on how they behave as functions of the single-particle energy eigenvalues (diagonal matrix elements) and their differences (off-diagonal matrix elements). Section~\ref{sec:fluctuations} is devoted to the study of the eigenstate-to-eigenstate fluctuations of the diagonal matrix elements, and the variances (and the ratios thereof) of the diagonal and off-diagonal matrix elements. In Sec.~\ref{sec:distributions} we discuss the distributions. We contrast one-body observables that exhibit non-Gaussian distributions to those that exhibit Gaussian ones. A summary and discussion of our results is presented in Sec.~\ref{sec:conclusions}.

\section{Models and Observables} \label{sec:models}

We consider two quadratic models in a lattice with $V$ sites. The models are particle-number conserving and we only study the single-particle sector, so the particle statistics plays no role. The first model is the quadratic Sachdev-Ye-Kitaev model in the Dirac fermion formulation (in short, the Dirac SYK2 model). We construct its Hamiltonian as a random matrix drawn from the Gaussian orthogonal ensemble in the position basis,
\begin{equation} \label{eq_Hsyk2}
\hat{H}_\text{SYK2} = \sum_{i,j=1}^{V} A_{ij} \hat{c}_{i}^\dagger \hat{c}_{j}\,,
\end{equation} 
where the diagonal (off-diagonal) elements of the symmetric matrix $A$ are real random numbers that are normally distributed with zero mean and $2/V$ ($1/V$) variance. The operator $\hat{c}_{i}^\dagger$ ($\hat{c}_{i}$) creates (annihilates) a particle at site $i$. In the thermodynamic limit, the mean single-particle energy is $\langle \hat{H}_\text{SYK2} \rangle = \frac{1}{V} {\rm Tr} \{ \hat{H}_\text{SYK2} \} = 0$ and the variance is $\langle \hat{H}_\text{SYK2}^2 \rangle = 1$. Since the single-particle density of states forms a Wigner semicircle distribution, for which the ratio between the maximal value and the standard deviation is 2, we expect the single-particle eigenenergies of $\hat{H}_\text{SYK2}$ to approximately belong to the interval $E_\alpha \in [-2,2]$.

The second model is the 3D Anderson model on a cubic lattice,
\begin{equation}
\label{eq_HA}
\hat{H}_\text{A} = -t\sum_{\left<i,j\right>} \hat{c}_i^\dagger \hat{c}_j + \frac{W}{2}\sum_{i=1}^V \epsilon_i \hat{c}_i^\dagger \hat{c}_i \,,
\end{equation}
where $t\equiv 1$ is the hopping integral between nearest neighbor sites (defined as $\langle i,j\rangle$), $\{\epsilon_i\}$ are independent and identically distributed random numbers drawn from a uniform distribution in an interval $\left[-1,1\right]$, and $W$ is the disorder strength. We assume periodic boundary conditions. The indices in Eq.~(\ref{eq_HA}) are defined as $i=x+\left(y-1\right)L+\left(z-1\right)L^2$ with $\left(x,y,z\right)$ standing for the Cartesian coordinates of sites, each belonging to the set $[1,...,L]$ with $L=V^{1/3}$.

The localization transition in the 3D Anderson model occurs at $W_{\rm c} \approx 16.5$~\cite{slevin_ohtsuki_18}. Unless otherwise specified, we focus on $W=1$, which is well below the localization transition, so that we have a quantum-chaotic quadratic model~\cite{lydzba_rigol_21}. (Results for the 3D Anderson insulator at $W\gg 16.5$ are also briefly discussed, and reported in Appendix~\ref{sec:breakdown}.) As for the $\hat{H}_\text{SYK2}$, the mean single-particle energy in the thermodynamic limit is $\langle \hat H_{\rm A} \rangle = 0$, and the variance is a constant that does not scale with the volume of the system (specifically, $\langle \hat H_{\rm A}^2 \rangle = 6 + W^2/12$~\cite{suntajs_prosen_21}). At weak disorder, the single-particle eigenenergies lie to a good approximation within the free fermion bandwidth, $E_\alpha \in [-6,6]$. The single-particle density of states evolves with increasing $W$ from the 3D free fermion distribution at $W=0$ towards the box distribution at $W = \infty$, see, e.g.,~Ref.~\cite{markos_06}. The 3D Anderson model has been widely studied in the literature, in particular from the perspective of its transport properties, spectrum fluctuations, and the structure of its single-particle eigenfunctions (see, e.g., Refs.~\cite{kramer_mackinnon_93, markos_06, evers_mirlin_08, suntajs_prosen_21} for reviews).

In the single-particle sector of the Hilbert space, the Hamiltonians~(\ref{eq_Hsyk2}) and~(\ref{eq_HA}) can be written in a general form $\hat{H}=\sum_{i,j=1}^{V}H_{ij}|i\rangle \langle j|$, where $H_{ij} = \langle i|\hat{H}|j\rangle$ and $\{|i\rangle\}$ is the single-particle site-occupation basis. The $V\times V$ matrix ${\bf H}$, with matrix elements $H_{ij}$, is diagonalized by a unitary $V\times V$ matrix ${\bf U}$, with matrix elements $U_{i\alpha}=\langle i|\alpha\rangle$. The resulting diagonal matrix ${\bf D}={\bf U}^\dagger {\bf H} {\bf U}$ has matrix elements $D_{\alpha\beta}=E_\alpha\delta_{\alpha\beta}$.

Note that we refer to the models under consideration as quantum-chaotic quadratic since the statistical properties of their single-particle spectra agree with the RMT predictions~\cite{altshuler_shklovskii_86, altshuler_zharekeshev_88, shklovskii_shapiro_93, hofstetter_schreiber_93, sierant_delande_20, suntajs_prosen_21}. This type of quantum chaos is sometimes referred to as {\it single-particle} quantum chaos. In contrast to previous studies of these models, our focus is on the expectation values of observables $\hat O$ in the single-particle energy eigenstates $\{|\alpha\rangle\}$ of the Hamiltonians in Eqs.~(\ref{eq_Hsyk2}) and~(\ref{eq_HA}). 

Throughout the presentation, observables $\underline{\hat O}$ (i.e., using underlined letters) are traceless 
\begin{equation} \label{def_trac}
\langle \underline{\hat O} \rangle = \frac{1}{V} {\rm Tr} \{ \underline{\hat O} \} = 0,
\end{equation}
and normalized
\begin{equation} \label{def_norm}
||\underline{\hat O}||^2 \equiv \frac{1}{V}{\rm Tr}\{ \underline{\hat O}^2 \}=1,
\end{equation}
namely, they have a unit Hilbert-Schmidt norm (also known as the Frobenius norm). The normalized counterparts of observables are important for the comparison of the numerical results reported here to those for quantum-chaotic interacting systems, because the ETH ansatz in Eq.~\eqref{def_eth_ansatz} is written having normalized observables in mind~\cite{leblond_mallayya_19, mierzejewski_vidmar_20}. In contrast, we label the experimentally measured one-body observables using letters that are not underlined. Most of them, such as the ones in Eqs.~(\ref{def_ni})-(\ref{def_m0}), have a unit Hilbert-Schmidt norm in the many-body Hilbert space. This is not the case in the single-particle Hilbert space.

We focus on the following observables:
(i) The site occupation
\begin{equation} \label{def_ni}
\hat n_i = \hat{c}_{i}^{\dagger}\hat{c}^{}_{i}, \quad
\underline{\hat{n}}_i = \frac{1}{\sqrt{V-1}}(V \hat n_i  - 1)\,.
\end{equation}
Without loss of generality, we fix $i=1$ and replace $\hat n_1 \to \hat n$ and $\underline{\hat n}_1 \to \underline{\hat n}$ to simplify the notation. (ii) The next-nearest neighbor correlation 
\begin{equation} \label{def_hij}
\hat{h}_{ij} = \hat{c}_{i}^{\dagger}\hat{c}^{}_{j}+\hat{c}_{j}^{\dagger}\hat{c}^{}_{i}, \qquad
\underline{\hat{h}}_{ij} = \sqrt{\frac{V}{2}}\hat{h}_{ij}\,.
\end{equation}
We fix $i=1$, coordinates (1,1,1), and $j=2+L$, coordinates (2,2,1), such that the correlations are measured along the diagonal within a plane, and replace $\hat h_{1,2+L} \to \hat h$ and $\underline{\hat h}_{1,2+L} \to \underline{\hat h}$.
(iii) The occupation of the zero quasi-momentum state
\begin{equation} \label{def_m0}
\hat{m}_{0} = \frac{1}{V} \sum_{i,j=1}^{V} \hat{c}_{i}^{\dagger}\hat{c}_{j}, \qquad
\underline{\hat m}_{0} = \frac{1}{\sqrt{V-1}}\left(V\hat{m}_{0}-1\right)\,.
\end{equation}

We note that for local Hamiltonians, such as the 3D Anderson model, the site occupation and the next-nearest neighbor correlation are local operators, while the occupation of the zero quasi-momentum state is nonlocal. We also highlight that the experimentally measured observables $\hat{n}$, $\hat{h}$, $\hat{m}_0$ and their normalized versions $\underline{\hat{n}}$, $\underline{\hat{h}}$, $\underline{\hat{m}}_0$ differ by multiplicative factors that depend on the number of lattice sites $V$. In addition, we note that the expectation values of these observables in single-particle energy eigenstates are expected to vanish when $V\rightarrow\infty$, because the average site occupation vanishes as $1/V$.

We study another local operator that does not suffer from the latter drawback, namely, the ``kinetic energy'' operator. Having a cubic lattice in mind, it can be written in the following form,
\begin{equation}
\label{def_T}
    \hat{T}=-\sum_{\langle i,j\rangle} \left(\hat{c}_{i}^{\dagger}\hat{c}^{}_{j} + \hat{c}_{j}^{\dagger}\hat{c}^{}_{i} \right) \,, \qquad \underline{\hat{T}}=\frac{1}{\sqrt{6}}\hat{T}\,,
\end{equation}
where $\langle i,j\rangle$ stands for nearest neighbor sites. We note that $\hat{T}$ and $\underline{\hat{T}}$ differ by a system-size independent multiplicative factor, and their expectation values in single-particle energy eigenstates do not need to vanish in the limit $V\rightarrow\infty$.

We stress that we only study one-body observables. In the single-particle sector (in systems with a particle number conservation), the matrix elements of multi-body observables can either be written in terms of matrix elements of one-body observables or they vanish.

For brevity, we will denote the matrix elements of observables in single-particle energy eigenstates as
\begin{equation}
\label{def_matele}
O_{\alpha\beta} \equiv \langle \alpha | \hat O |\beta \rangle.
\end{equation}
In what follows we drop the``single-particle'' prefix as our focus is on the single-particle sector, while we keep the ``many-body'' prefix when many-body states are considered.

\section{Structure of matrix elements} \label{sec:structure}

\subsection{Diagonal matrix elements}

We first study the diagonal matrix elements of observables in all eigenstates of the 3D Anderson and Dirac SYK2 models. Our goal is to unveil how they behave as functions of the energy when increasing the number of lattice sites. Since we are dealing with a single particle in an increasingly large lattice, we multiply the matrix elements of $\hat n,\,\hat h,\,$ and $\hat m_0$ by $V$ to ensure that the scaled matrix elements are of order 1. The quantitative analysis of the eigenstate-to-eigenstate fluctuations and variances of the diagonal matrix elements is carried out in Sec.~\ref{sec:fluctuations} (for the normalized observables).

%%%%%%%%%%%%%%%%%%%%%%%%%%%%%%%% FIGURE 1 %%%%%%%%%%%
\begin{figure}[!t]
\centering
\includegraphics[width=0.98\columnwidth]{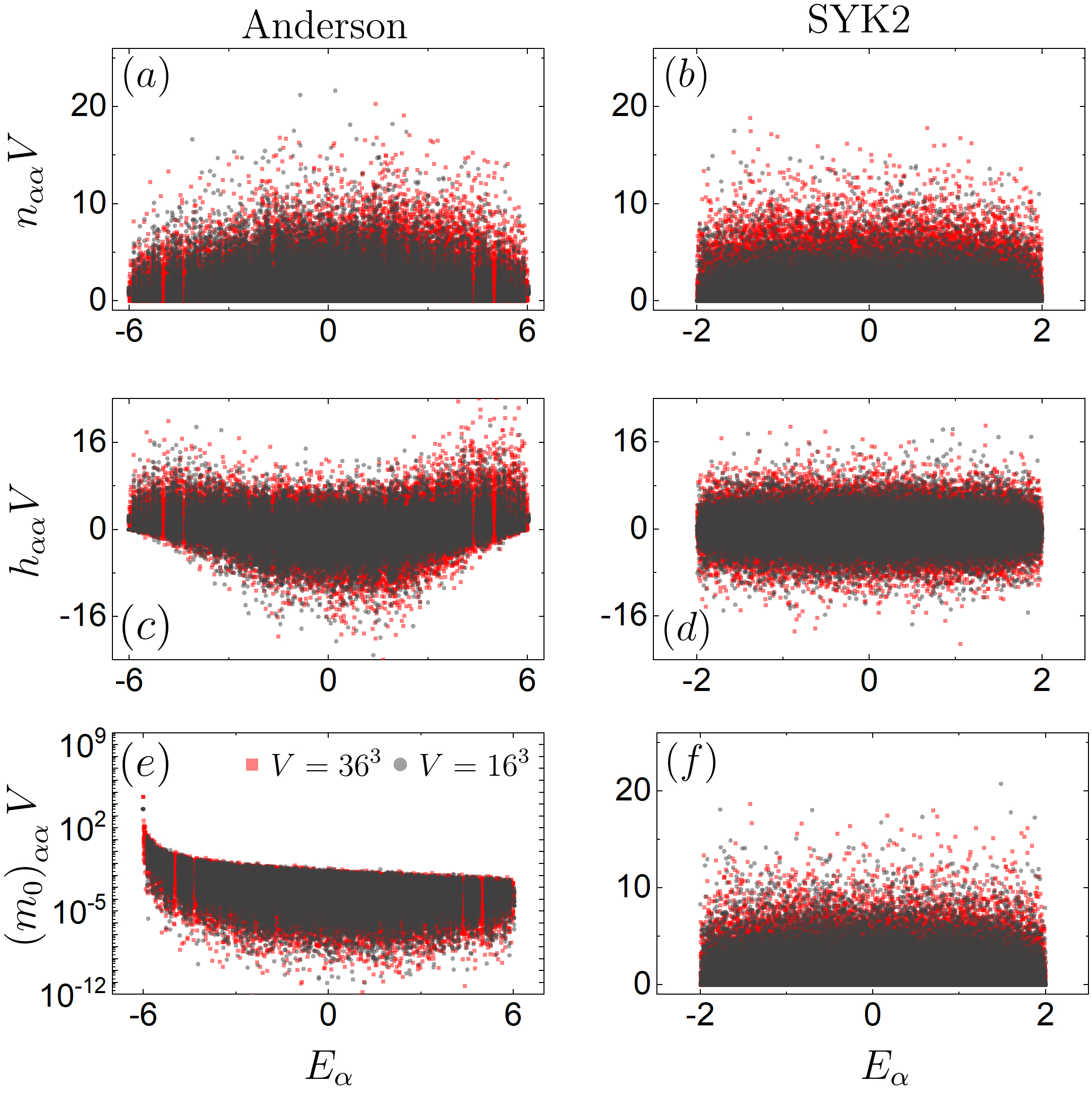}
\caption{Diagonal matrix elements of the observables (a),(b) $\hat n$, (c),(d) $\hat h$, and (e),(f) $\hat m_0$ as functions of the eigenenergies $E_\alpha$. Results are shown for the 3D Anderson model (left column) and the Dirac SYK2 model (right column). Each panel shows results for two system sizes $V=16^3$ and $36^3$ with $20$ (for $V=16^3$) and $3$ (for $V=36^3$) different Hamiltonian realizations. The points are half-transparent, which means that the darker the color, the more overlapping points.}\label{fig1}
\end{figure}

In Fig.~\ref{fig1}, we show results for the 3D Anderson model (left column) and for the Dirac SYK2 model (right column). For the Dirac SYK2 model, the diagonal matrix elements are structureless and the eigenstate-to-eigenstate fluctuations of $O_{\alpha\alpha}V$ do not significantly change with increasing system size (see Sec.~\ref{sec:fluctuations}). This signals that the fluctuations of the traditional and normalized observables, like their expectation values, vanish in the thermodynamic limit. Another interesting aspect of the Dirac SYK2 model is the striking similarity between the matrix elements of $\hat n V$ and $\hat m_0 V$, see Figs.~\ref{fig1}(b) and~\ref{fig1}(f). This can be easily understood because those operators are occupation operators in two different spaces, and the base kets of those two spaces can be equivalently considered as random vectors in the eigenbasis of $\hat H_{\rm SYK2}$~\cite{dalessio_kafri_16}.

The diagonal matrix elements of observables in the 3D Anderson model, in contrast, may exhibit nontrivial structures. For example, $h_{\alpha\alpha}V$ has a quadratic structure, see Fig.~\ref{fig1}(c), which may be understood~\cite{mierzejewski_vidmar_20} as being a consequence of a nonzero projection of $\hat h V$ onto the square of the Hamiltonian. Particularly interesting is the structure of $(m_0)_{\alpha\alpha}V$ in Fig.~\ref{fig1}(e). The ground-state matrix element dominates the spectrum, i.e., its value is several orders larger than these of excited-states matrix elements [note the logarithmic scale in Fig.~\ref{fig1}(e)], and $(m_0)_{\alpha\alpha}V$ appears to be an exponentially decaying function of a single-particle eigenenergy $E_\alpha$. This is a consequence of the proximity of the $W=1$ case considered to the translationally-invariant free fermion point at $W=0$. The large value of the ground-state matrix element will impact the analysis of fluctuations and distributions of normalized observables in Secs.~\ref{sec:fluctuations} and~\ref{sec:distributions}, respectively.

\begin{figure}[!t]
\centering
\includegraphics[width=0.98\columnwidth]{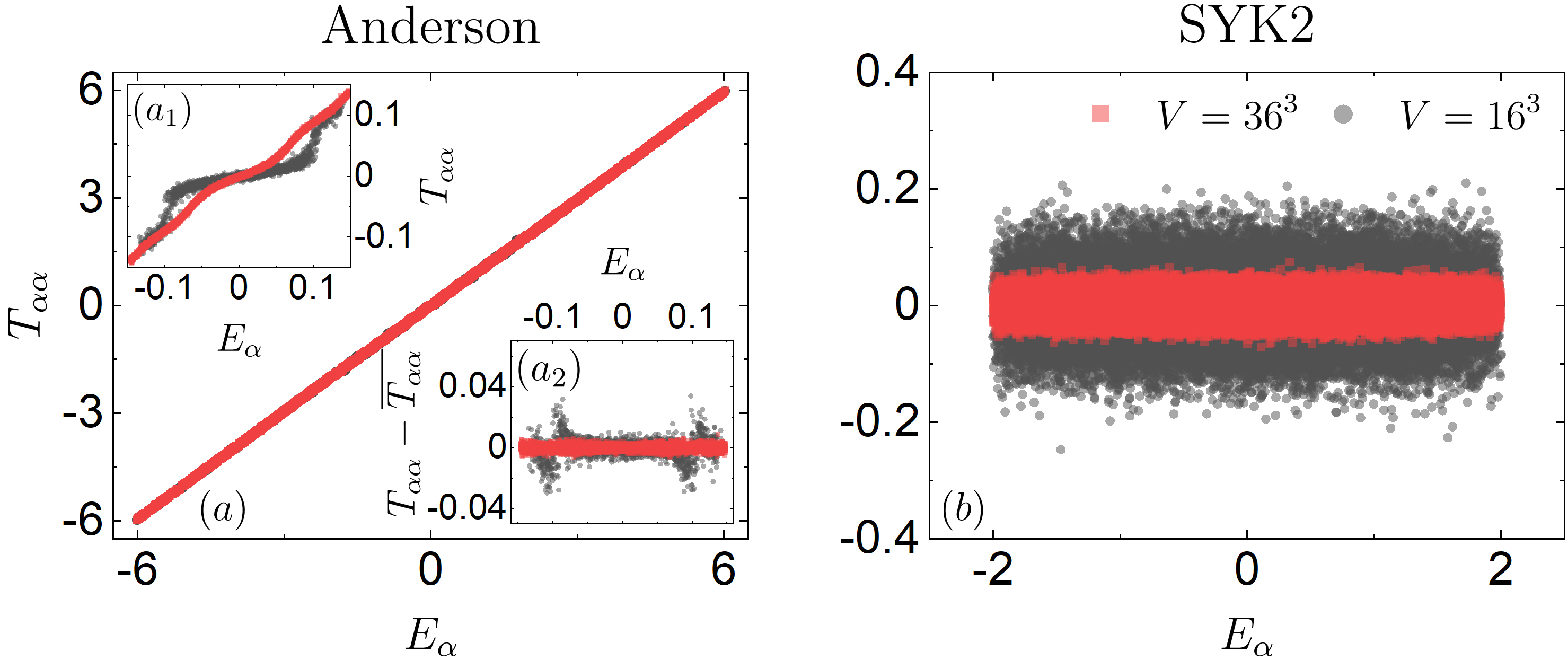}
\caption{Diagonal matrix elements of $\hat{T}$ vs.~the eigenenergies $E_\alpha$. Results are shown for (a) the 3D Anderson model and (b) the Dirac SYK2 model. Each panel shows results for two system sizes $V=16^3$ and $36^3$ with $10$ (for $V=16^3$) and $3$ (for $V=36^3$) different Hamiltonian realizations. The inset ($\text{a}_1$) is a close-up of the main panel including $200$ (for $V=16^3$) and $2000$ (for $V=36^3$) diagonal matrix elements from the center of the spectrum. The inset ($\text{a}_2$) shows the same matrix elements as ($\text{a}_1$) after the moving average $\overline{T_{\alpha\alpha}}$ is removed. For the target eigenstate $\alpha$, the moving average $\overline{T_{\alpha\alpha}}$ is computed using the matrix elements of $20$ closest eigenstates. The points used to show the diagonal matrix elements are half-transparent.}
\label{fig2}
\end{figure}

In Fig.~\ref{fig2}, we show the diagonal matrix elements of the kinetic energy $\hat T$ from Eq.~(\ref{def_T}). They are linearly dependent on the eigenenergies in the 3D Anderson model [Fig.~\ref{fig2}(a)], while (as expected) there is no structure in the Dirac SYK2 model [Fig.~\ref{fig2}(b)]. The linear dependence in the 3D Anderson model originates from the nonzero projection of $\hat T$ onto the Hamiltonian $H_{\rm A}$ from Eq.~(\ref{eq_HA}), which is the sum of $\hat T$ and onsite disorder. A detailed inspection of $T_{\alpha\alpha}$ as a function of $E_\alpha$ in small systems, see Fig.~\ref{fig2}($\text{a}_1$), reveals a fine structure beyond this linear dependence, which becomes less pronounced with increasing system size. When studying the variances of the diagonal matrix elements in the next sections, we subtract the moving average, $T_{\alpha\alpha} \to T_{\alpha\alpha} - \overline{T_{\alpha\alpha}}$, where $\overline{T_{\alpha\alpha}}$ is the arithmetic mean of closest diagonal matrix elements about $\alpha$. The relation between $T_{\alpha\alpha}$ and $E_\alpha$ after the subtraction of the moving average is shown in Fig.~\ref{fig2}($\text{a}_2$). Note that the eigenstate-to-eigenstate fluctuations of $T_{\alpha\alpha}$, both in the 3D Anderson [Fig.~\ref{fig2}(a)] and Dirac SYK2 [Fig.~\ref{fig2}(b)] models, decrease with increasing system size.

%%%%%%%%%%%%%%%%%%%%%%%%%%%%%%%% FIGURE 2 %%%%%%%%%%%

\subsection{Off-diagonal matrix elements} \label{sec:structure_offdiag}

We next focus on the structure of the square of the off-diagonal matrix elements. They are multiplied by an appropriate power of $V$, such that the scaled coarse-grained matrix elements are $V$-independent, i.e., we study $O_{\alpha\beta}^2V^2$ where $\hat O = \hat n, \hat h$ and $\hat m_0$, and $T_{\alpha\beta}^2V$. We restrict the pairs of eigenstates $|\alpha\rangle, |\beta\rangle$ to a narrow energy window $\Delta$ around a target energy $\bar E_{\rm tar}$, $|(E_\alpha+E_\beta)/2 - \bar E_{\rm tar}|<\Delta/2$. We take the target energy to be the mean energy of the entire spectrum, and the width to be $\Delta=(E_V-E_1)/100$ ($E_1$ and $E_V$ are the ground state and the highest excited state energies, respectively). Even though $\bar E_{\rm tar}$ is very close to zero in finite systems (and $\bar E_{\rm tar} = 0$ in the thermodynamic limit), we calculate both $\bar E_{\rm tar}$ and $\Delta$ numerically for each Hamiltonian realization.

%%%%%%%%%%%%%%%%%%%%%%%%%%%%%%%% FIGURE 3 %%%%%
\begin{figure}[!t]
\centering
\includegraphics[width=0.98\columnwidth]{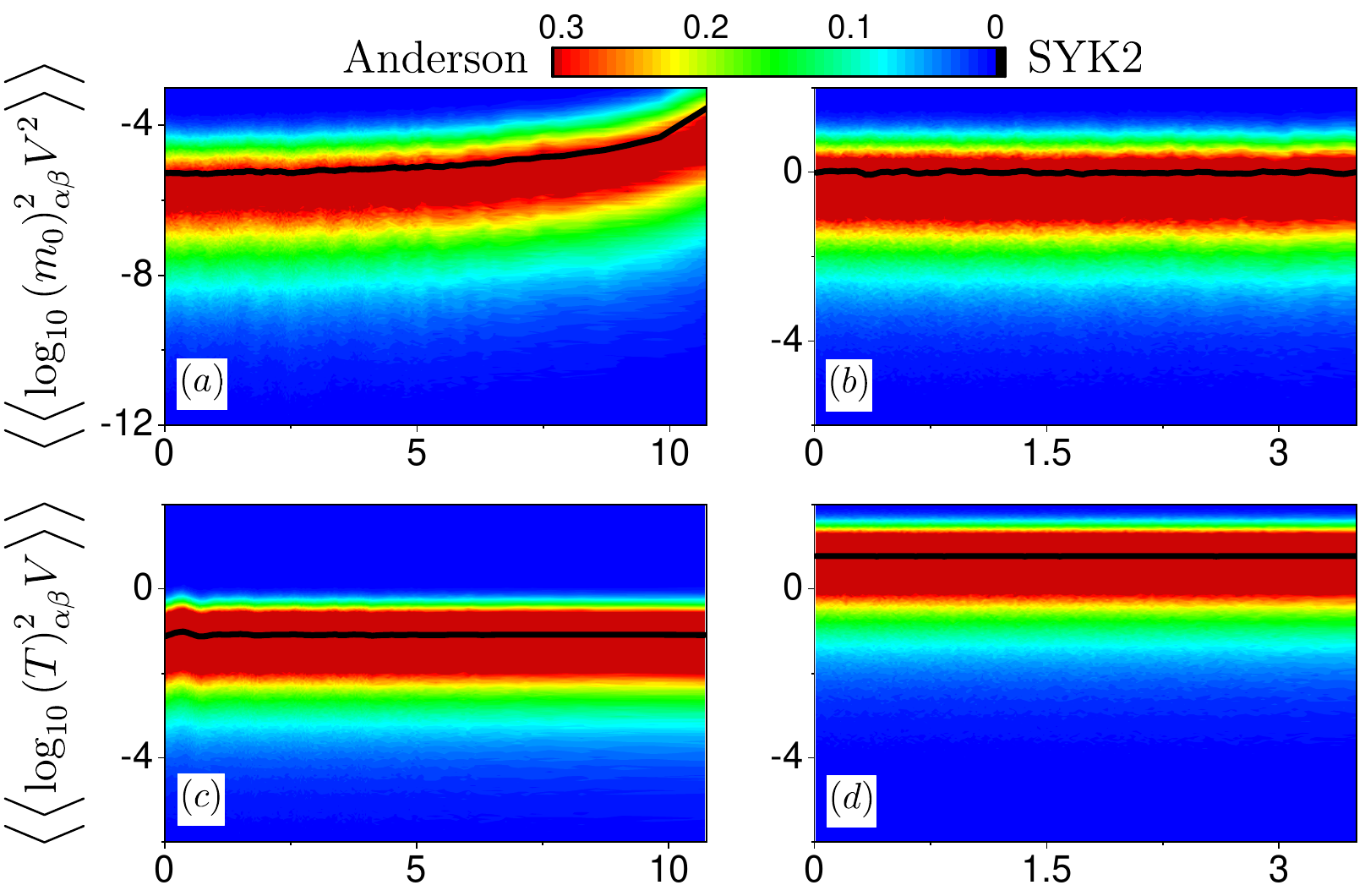}
\caption{Density plots of the off-diagonal matrix elements of the observables (a),(b) $\hat m_0$ and (c),(d) $\hat T$ as functions of the energy difference $\omega = |E_\alpha - E_\beta|$. We consider $V=20^3$ for (a),(b) and $V=22^3$ for (c),(d). Results for the 3D Anderson model (left column) and the Dirac SYK2 model (right column) are shown for pairs of eigenstates that belong to the target energy window, as explained in the text. For a given disorder realization, we discretize both axes and calculate $\log_{10}[(m_0)_{\alpha\beta}^2 V^2]$ and $\log_{10}(T_{\alpha\beta}^2 V)$ in each bin. We then average these values over 20 Hamiltonian realizations, yielding $\langle\langle \log_{10} (m_0)_{\alpha\beta}^2 V^2 \rangle\rangle$ and $\langle\langle \log_{10} T_{\alpha\beta}^2 V \rangle\rangle$. The black lines denote the moving averages of the results $\langle\langle \log_{10} \overline{ (m_0)_{\alpha\beta}^2} V^2\rangle\rangle$ and $\langle\langle \log_{10} \overline{ T_{\alpha\beta}^2} V\rangle\rangle$ (see text for details).}
\label{fig3}
\end{figure}

Figure~\ref{fig3} shows the density plots of the logarithms of off-diagonal matrix elements, $\log_{10}[(m_0)_{\alpha\beta}^2 V^2]$ and $\log_{10}(T_{\alpha\beta}^2 V)$, as functions of the energy difference $\omega = |E_\alpha - E_\beta|$. To smooth out fluctuations, we carry out an average over 20 different Hamiltonian realizations, and denote the realization averaged results as $\langle\langle ... \rangle\rangle$. The corresponding density plots for the site occupation $\hat n$ and the next-nearest neighbor correlation $\hat h$ are shown in Fig.~\ref{figA3} of Appendix~\ref{sec:offdiag_appendix}. Figures~\ref{fig3} and~\ref{figA3}, as well as a detailed inspection of individual off-diagonal matrix elements (not shown), demonstrates that the off-diagonal matrix elements are dense, i.e., there is no set comprising a considerable number of off-diagonal matrix elements that are zero (or below numerical precision). This is similar to what is observed in quantum-chaotic interacting systems.

The black lines in Fig.~\ref{fig3} show results obtained for moving averages as functions of $\omega$. Specifically, we order the scaled matrix elements in $\omega$, and divide them into $150$ $\omega$-bins. Next, we calculate the mean $\overline{O_{\alpha\beta}^2} V^\eta$ ($\eta = 1$~or~$2$) within each $\omega$-bin, and then average the logarithm of the latter over different Hamiltonian realizations, yielding $\langle\langle \log_{10}\overline{O_{\alpha\beta}^2} V^\eta \rangle\rangle$. For the Dirac SYK2 model (right panels in Fig.~\ref{fig3}), the results for the moving averages make apparent something that was already visible at the level of the density plots, namely, that the off-diagonal matrix elements are structureless. Note that $\langle\langle \log_{10} \overline{O_{\alpha\beta}^2} V^\eta \rangle\rangle \approx 0$ implies that $\overline{O_{\alpha\beta}^2} \approx 1/V^\eta$, i.e., the coarse-grained matrix elements are nearly identical throughout the spectrum. This is expected (after properly normalizing the observables) within the random matrix theory~\cite{dalessio_kafri_16}.

\begin{figure}[!t]
\centering
\includegraphics[width=0.98\columnwidth]{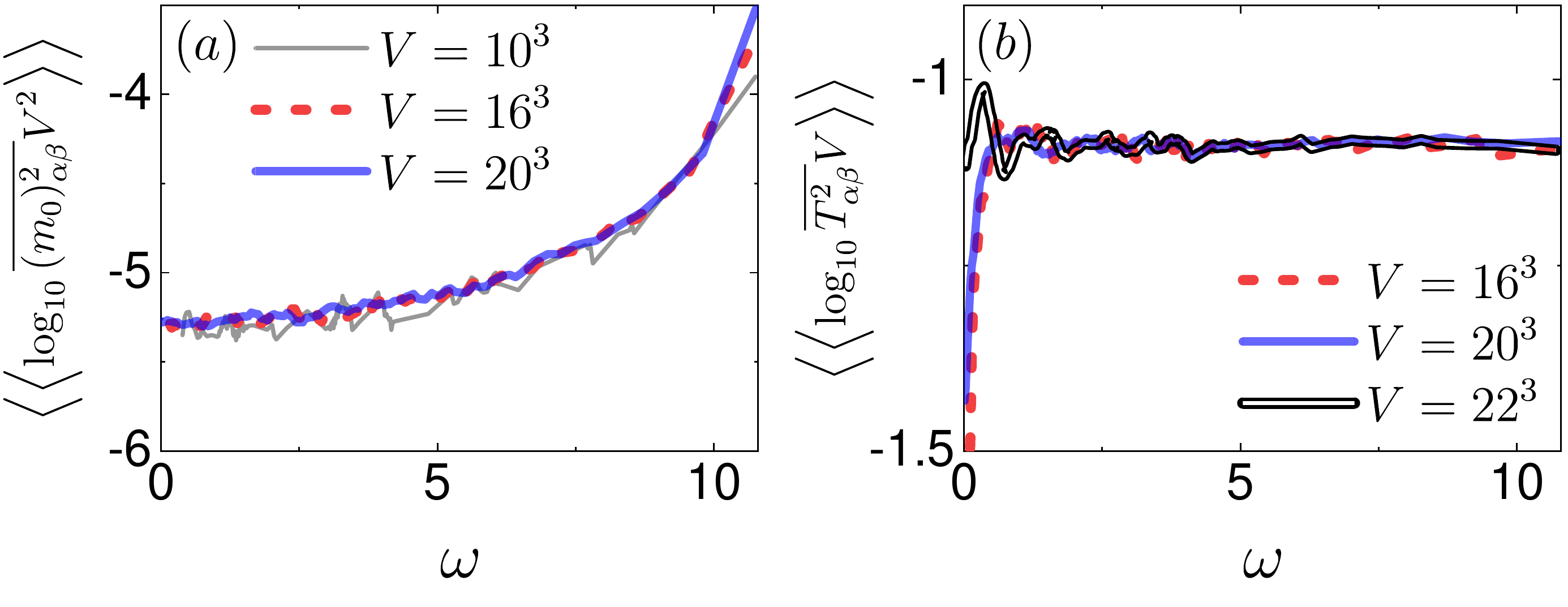}
\caption{Moving averages (a) $\langle\langle \log_{10}\overline{(m_0)_{\alpha\beta}^2} V^2 \rangle\rangle$ and (b) $\langle\langle \log_{10}\overline{T_{\alpha\beta}^2} V \rangle\rangle$ versus $\omega$ for different system sizes in the 3D Anderson model (see text for details).}
\label{fig4}
\end{figure}

On the other hand, the results for the 3D Anderson model (left panels in Fig.~\ref{fig3}) may exhibit a structure. To highlight it, when present, we show the moving averages of observables $\hat m_0$ and $\hat T$ for different $V$ in Fig.~\ref{fig4}. Notice that the results for different system sizes, away from the $\omega\rightarrow 0$ and $\omega\rightarrow |E_V-E_1|$ limits, exhibit excellent data collapse. For $\hat m_0$, see Fig.~\ref{fig4}(a), the coarse-grained values of the off-diagonal matrix elements increase at high $\omega$, indicating that there are large matrix elements between energy eigenstates in the lowest and highest part of the energy spectrum (which, in single-particle systems, has a bandwidth independent of $V$). For $\hat T$, see Fig.~\ref{fig4}(b), the coarse-grained values of the off-diagonal matrix elements are nearly $\omega$-independent at high $\omega$. In quantum-chaotic interacting systems, in contrast, the matrix elements become exponentially small at high $\omega$~\cite{dalessio_kafri_16}.

%%%%%%%%%%%%%%%%%%%%%%%%%%%%%%%% FIGURE 4 %%%%%%

As a technical remark we note that we excluded the matrix elements between excited states and the ground state from the results reported for the off-diagonal matrix elements of the quasi-momentum occupation $\hat m_0$. The reason is that those matrix elements are several orders of magnitude larger than the others, and give rise to pronounced fluctuations of the moving average $\langle\langle\log_{10}\overline{(m_0)_{\alpha\beta}^2} V^2\rangle\rangle$. A similar exceptionally large value was observed for the ground-state diagonal matrix element of $\hat m_0$ in Fig.~\ref{fig1}(e).

\section{Fluctuations of matrix elements} \label{sec:fluctuations}

\subsection{Eigenstate-to-eigenstate fluctuations} \label{sec:eigtoeig}

We now turn our attention to the quantitative analysis of the fluctuations of the matrix elements. We first study the eigenstate-to-eigenstate fluctuations of the diagonal matrix elements of normalized observables: $\delta \underline{O}_\alpha = \underline{O}_{\alpha,\alpha} - \underline{O}_{\alpha-1,\alpha-1}$. We calculate the average of the absolute values of these differences
\begin{equation} \label{def_dO}
\delta \underline{O}_\text{av}=||\Lambda||^{-1} \sum_{\ket{\alpha}\in\Lambda} |\delta \underline{O}_\alpha| \,,
\end{equation}
where $\Lambda$ is a set of states $|\alpha\rangle$ that comprise $80\%$ of eigenstates in the middle of the spectrum, i.e., $||\Lambda||=0.8V$. We also calculate the maximal difference as
\begin{equation} \label{dO_max}
\delta \underline{O}_\text{max} = \text{max}_{\ket{\alpha}\in\Lambda}|\delta \underline{O}_\alpha| \,.
\end{equation}
In practice, we first calculate $\delta \underline{O}_\text{av}$ and $\delta \underline{O}_\text{max}$ for a single Hamiltonian realization from $80\%$ ($200$) of eigenstates in the middle of the spectrum for $\underline{\hat{n}}$, $\underline{\hat{h}}$, $\underline{\hat{m}}_0$ ($\underline{\hat{T}}$), and then average the results over $100$ ($20$) Hamiltonian realizations for $V<28^3$ ($V\ge 28^3)$. We denote the latter averages as $\langle\langle \delta \underline{O}_\text{av} \rangle\rangle$ and $\langle\langle \delta \underline{O}_\text{max} \rangle\rangle$, respectively. The analysis of the fluctuations between Hamiltonian realizations, and their scaling with the number of lattice sites, is presented in Appendix~\ref{sec:realizations}.

The eigenstate-to-eigenstate fluctuations were proposed as a simple measure to test the eigenstate thermalization in quantum-chaotic interacting systems~\cite{kim_ikeda_14}. A particularly strong indicator of the eigenstate thermalization is the vanishing of maximal differences $\delta \underline{O}_\text{max}$~(\ref{dO_max}) with increasing system size. In quantum-chaotic interacting systems, numerical studies of several models showed polynomial decay of $\delta \underline{O}_\text{max}$ with the Hilbert space dimension~\cite{mondaini_fratus_16, luitz_16, jansen_stolpp_19}. For the models considered here, since the Hilbert space dimension is $V$, an analogous scaling would imply a decay $\propto 1/V^\zeta$ with $\zeta > 0$.

%%%%%%%%%%%%%%%%%%%%%%%%%%%%%%%% FIGURE 5 %%%%%%
\begin{figure}[!t]
\centering
\includegraphics[width=0.98\columnwidth]{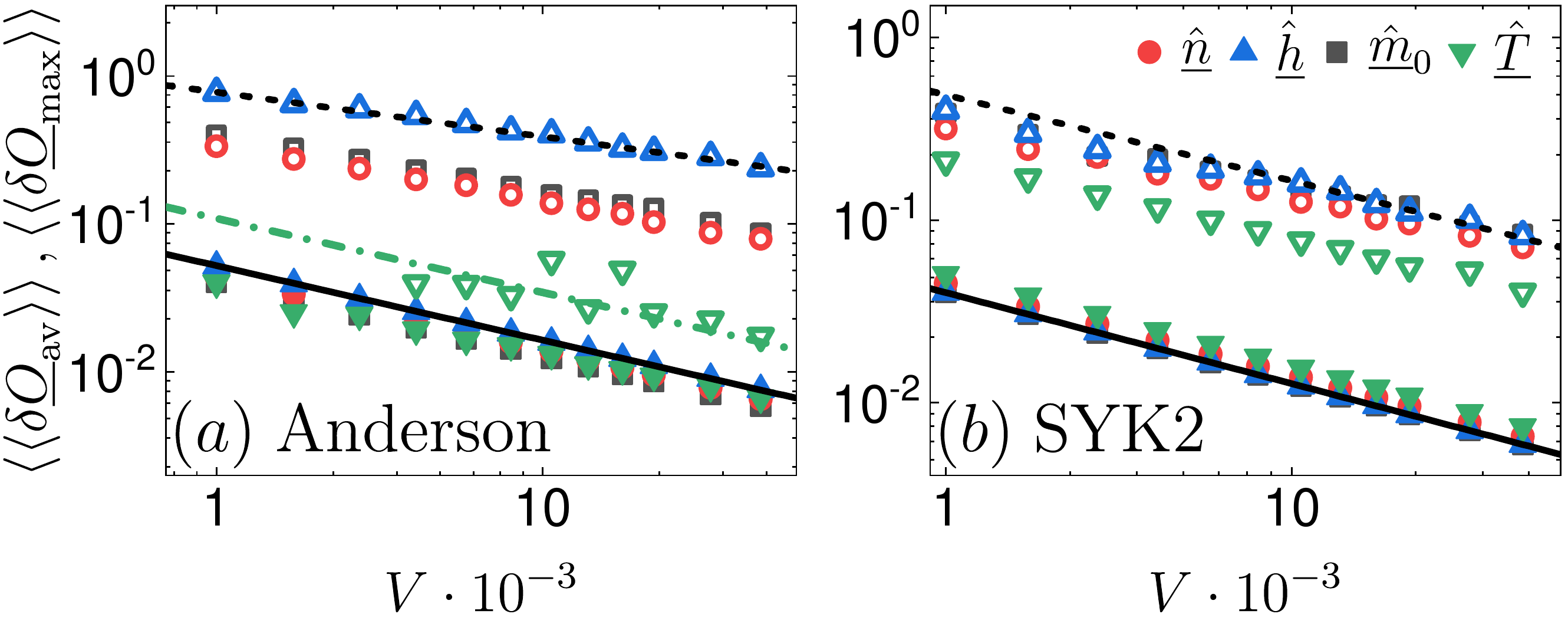}
\caption{Eigenstate-to-eigenstate fluctuations in (a) the 3D Anderson model, and (b) the Dirac SYK2 model. Filled symbols: $\langle\langle \delta \underline{O}_\text{av} \rangle\rangle$ versus $V$, see Eq.~(\ref{def_dO}). Open symbols: $\langle\langle \delta \underline{O}_\text{max} \rangle\rangle$ versus $V$, see Eq.~(\ref{dO_max}). The matrix elements of $\underline{\hat m}_0$ in (a) are multiplied by a constant $\gamma=420$, as explained in the text. The matrix elements of $\underline{\hat T}$ in (a) are multiplied by $8$ for clarity. All fluctuations are calculated from $80\%$ of the spectrum about the mean energy. The exception is $\langle\langle \delta \underline{T}_\text{max} \rangle\rangle$ in the 3D Anderson model, for which we only include 200 eigenstates about the mean energy (see also the main text). An averaging over $100$ (20) Hamiltonian realizations is carried out for $V<28^3$ ($V \ge 28^3$). The solid and dashed lines show fits of $a/V^\zeta$ to $\left( \underline{m}_0 \right)_{\alpha\alpha}$ for the largest five system sizes. We find $\zeta = 0.50$ for $\langle\langle \delta (\underline{m}_0) _\text{av}\rangle\rangle$ in both models (solid lines), and $\zeta=0.30$ (0.47) for $\langle\langle \delta (\underline{m}_0) _\text{max}\rangle\rangle$ in the 3D Anderson (Dirac SYK2) model (dashed lines). The dashed-dotted line following the results for $\langle\langle \delta \underline{T}_\text{max} \rangle\rangle$ in the 3D Anderson model shows $a/V^{0.5}$ and serves as a guide to the eye.}
\label{fig5}
\end{figure}

Results for $\langle\langle \delta \underline{O}_\text{av} \rangle\rangle$ and $\langle\langle \delta \underline{O}_\text{max} \rangle\rangle$ for the 3D Anderson model and the Dirac SYK2 model are presented in Fig.~\ref{fig5}. We indeed find power-law decays with $V$. Specifically, we find that the average $\langle\langle \delta \underline{O}_\text{av} \rangle\rangle$ is $\propto 1/V^{0.50}$ for all observables in both models, see the solid lines in Fig.~\ref{fig5}. The maximum $\langle\langle\delta \underline{O}_\text{max}\rangle\rangle$ is also $\propto 1/V^\zeta$, but with a slightly smaller $\zeta$, as shown by the dashed lines in Fig.~\ref{fig5} for the case of the quasi-momentum occupation $\langle\langle \delta ( \underline{m}_0 )_\text{max}\rangle\rangle$. For $\underline{\hat T}$ in the 3D Anderson model, $\langle\langle\delta \underline{O}_\text{max}\rangle\rangle$ exhibits large fluctuations but it is still consistent with a power law decay $\propto 1/V^{0.5}$ (shown as a dashed-dotted line). We also fitted the results for the other observables to $a/V^\zeta$ (not shown), and obtained $0.3 < \zeta < 0.5$. The fact that the values of $\zeta$ approach 0.5 with increasing system size in the SYK2 model [see the fit of $\langle\langle\delta \underline{O}_\text{max}\rangle\rangle$ in Fig.~\ref{fig5}(b)] suggests that the deviations of $\zeta$ from 0.5 found in our numerical calculations may result from finite-size effects.

We note that in Fig.~\ref{fig5}(a) some operators are multiplied by a global constant for clarity. Such a multiplication does not modify the exponent $\zeta$ in the power-law scalings of the eigenstate-to-eigenstate fluctuations [lines in Fig.~\ref{fig5}(a)]. Specifically, the matrix elements of the kinetic energy $\underline{\hat{T}}$ in the 3D Anderson model are multiplied by a global constant 8, and the matrix elements of the quasi-momentum occupation $\underline{\hat m}_0$ are multiplied by a global constant $\gamma$ (defined below). In the latter case, such a multiplication is convenient due to the exceptionally large ground-state matrix element $\left(\underline{m}_0\right)_{11}$ as shown in Fig.~\ref{fig1}(e), which, by the virtue of the Hilbert-Schmidt norm~(\ref{def_norm}), strongly reduces the values of other matrix elements. The specific value of $\gamma$ chosen is the ratio between the standard deviation of the diagonal matrix elements of $\underline{\hat{n}}$ and of $\underline{\hat{m}}_0$ in the small energy window comprising $200$ eigenenergies around the mean energy of the entire energy spectrum. We find $\gamma$ to be between $400$ and $450$ for systems with $V\leq 36^3$, and choose $\gamma=420$ for the results shown in Fig.~\ref{fig5}(a). (Similar multiplications are carried out in Figs.~\ref{fig6}-\ref{fig8}.)

Another technical remark is that the calculation of $\langle\langle \delta \underline{T}_\text{max} \rangle\rangle$ in the 3D Anderson model is sensitive to the fine structure of $\underline{T}_{\alpha\alpha}$ beyond the linear dependence of $\underline{T}_{\alpha\alpha}$ on $E_\alpha$ [see Fig. \ref{fig2}(a)]. To reduce the finite-size effects, we restrict the number of eigenstates to $200$, and we only consider systems with $V\ge 16^3$.

In Appendix~\ref{sec:breakdown}, we report results for the eigenstate-to-eigenstate fluctuations in the 3D Anderson model deep in the localized regime (at $W=35$). Even though the average fluctuations of $\underline{\hat{n}}$ and $\underline{\hat{h}}$ decay $\propto 1/V^\zeta$ with $\zeta\approx 0.5$, see Fig.~\ref{figA1}(a), the maximal fluctuations diverge $\propto V^\zeta$ with $\zeta\approx 0.5$, see Fig.~\ref{figA1}(b). The latter indicates the breakdown of the single-particle version of the ETH, which goes in parallel with the lack of quantum chaos in the energy spectrum.

\subsection{Variances of matrix elements}

Next we study the variances of matrix elements. The variance of the diagonal part is
\begin{equation} \label{def_variance_diag}
    \sigma^2_\text{diag}=||\Gamma||^{-1} \sum_{\ket{\alpha}\in\Gamma} \underline{O}^2_{\alpha\alpha}-
    \left(||\Gamma||^{-1}\sum_{\ket{\alpha}\in\Gamma} \underline{O}_{\alpha\alpha}\right)^2 \,,
\end{equation}
where $\Gamma$ is a set of 200 eigenstates ($||\Gamma|| = 200$) around the mean energy. Analogously, the variance of the off-diagonal part is
\begin{equation}
    \sigma^2_\text{off}=||\Gamma'||^{-1} \sum_{\substack{\ket{\alpha},\ket{\beta}\in\Gamma \\ \ket{\alpha}\neq\ket{\beta}}} \underline{O}^2_{\alpha\beta}
    -\left(||\Gamma'||^{-1} 
    \sum_{\substack{\ket{\alpha},\ket{\beta}\in\Gamma \\ \ket{\alpha}\neq\ket{\beta}}} \underline{O}_{\alpha\beta}\right)^2 \,,
\end{equation}
where $||\Gamma'|| = ||\Gamma||^2-||\Gamma||=39800$. We calculate $\sigma^2_\text{diag}$ and $\sigma^2_\text{off}$ for each Hamiltonian realization, and then average over $100$ different Hamiltonian realizations. We denote the latter averages as $\langle\langle \sigma_{\rm diag}^2 \rangle\rangle$ and $\langle\langle \sigma_{\rm off}^2 \rangle\rangle$, respectively.

%%%%%%%%%%%%%%%%%%%%%%%%%%%%%%%% FIGURE 6 %%%%%%
\begin{figure}[!t]
\centering
\includegraphics[width=0.85\columnwidth]{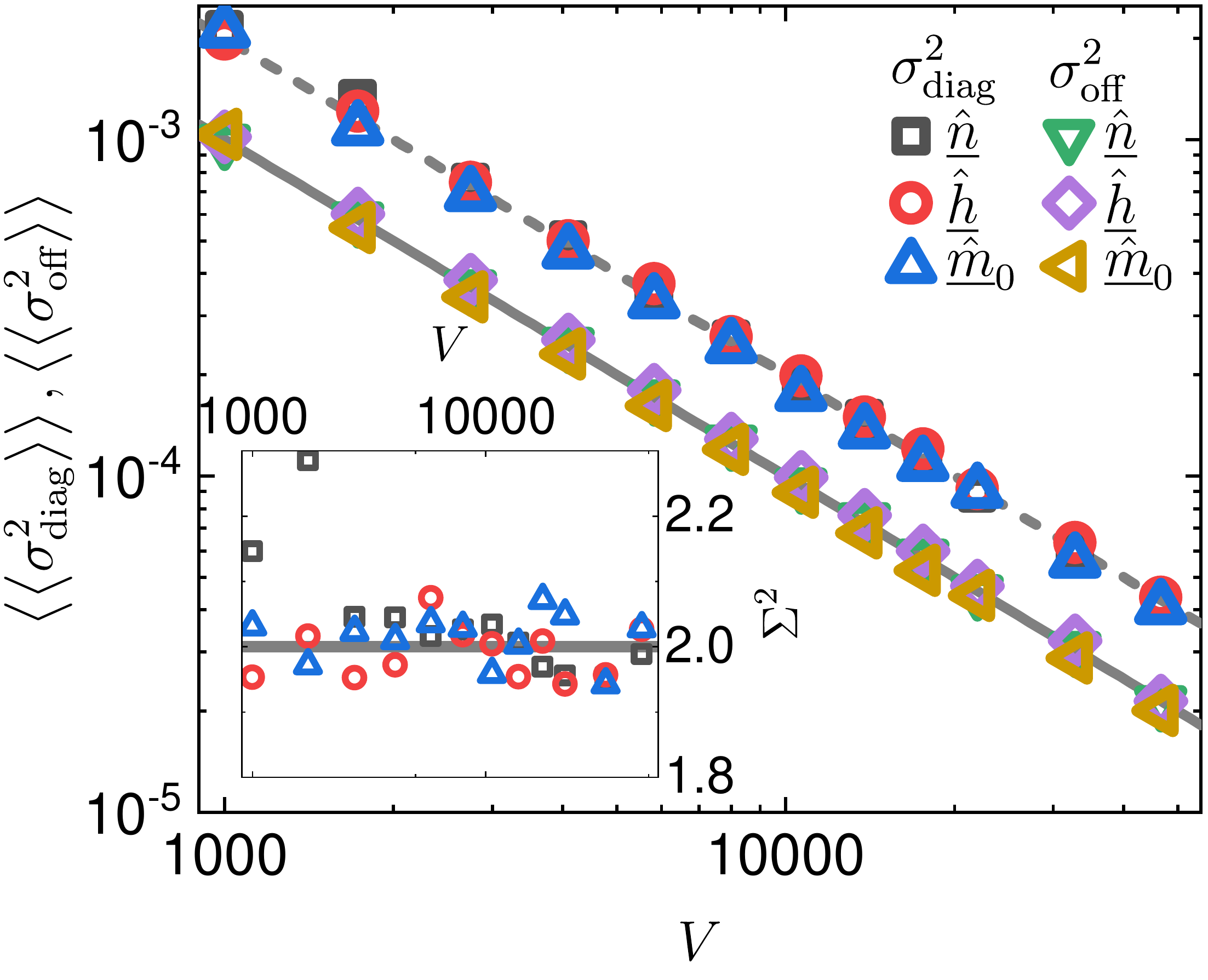}
\caption{Variances of the diagonal and off-diagonal matrix elements of observables in the 3D Anderson model. The matrix elements of $\underline{\hat{m}}_{0}$ are multiplied by $\gamma = 420$, as explained in Sec.~\ref{sec:eigtoeig}. The solid and dashed lines in the main panel show $1/V$ and $2/V$, respectively. The inset shows the ratio of variances $\Sigma^2$. For systems with $V>12^3$, the ratio lies within a small interval around $2$, i.e., $\Sigma^2 \in \left[1.9,2.1\right]$.}\label{fig6}
\end{figure}

Under the assumption that the eigenstates of a normalized observable ${\underline{\hat O}}$ are random vectors in the eigenbasis of the Hamiltonian under investigation~\cite{dalessio_kafri_16}, one can show that the associated variances in the Dirac SYK2 model are $\langle\langle \sigma_{\rm diag}^2 \rangle \rangle = 2/V$ and $\langle\langle \sigma_{\rm off}^2 \rangle \rangle = 1/V$, so that their ratio is 2.  Hence, in what follows we focus on the variances in the 3D Anderson model.

The variances of the matrix elements of $\underline{\hat n}$, $\underline{\hat h}$ and $\underline{\hat m}_0$ in the 3D Anderson model are shown in Fig.~\ref{fig6}. They exhibit the behavior advanced by the random matrix theory. Namely, both $\langle\langle \sigma_{\rm diag}^2 \rangle \rangle$ and $\langle\langle \sigma_{\rm off}^2 \rangle \rangle$ are $\propto 1/V$, as shown in the main panel of Fig.~\ref{fig6}. In quantum-chaotic interacting models, the scaling $\langle\langle \sigma_{\rm diag}^2 \rangle \rangle \propto 1/{\cal D}$ (where ${\cal D}$ is the dimension of the Hilbert space) is a hallmark of eigenstate thermalization for normalized observables~\cite{dalessio_kafri_16}. On the other hand, the scaling $\langle\langle \sigma_{\rm off}^2 \rangle \rangle \propto 1/{\cal D}$ is not necessarily the result of eigenstate thermalization, it can also be found in integrable models~\cite{leblond_mallayya_19}, and can be understood as a consequence of the normalization~(\ref{def_norm}) of observables.

We also note that the ratio of variances
\begin{equation} \label{def_ratio}
    \Sigma^2 = \frac{\langle\langle\sigma_{\rm diag}^2\rangle\rangle}{\langle\langle\sigma_{\rm off}^2\rangle\rangle} \,,
\end{equation}
shown in the inset in Fig.~\ref{fig6}, is very close to the RMT prediction $\Sigma^2 = 2$. This is again a hallmark of the validity of the ETH in quantum-chaotic interacting systems~\cite{mondaini_rigol_17, jansen_stolpp_19, schoenle_jansen_21}. We emphasize that we also find $\Sigma^2 = 2$ for the quasi-momentum occupation $\underline{\hat{m}}_0$, which exhibits a peculiar structure of the diagonal matrix elements, see Fig.~\ref{fig1}(e). This structure gives rise to much smaller variances of the matrix elements in the bulk of the spectrum as compared to those of the site occupation $\underline{\hat n}$ and next-nearest neighbor correlation $\underline{\hat h}$, but does not change the ratio of variances.

We emphasize that the ratio of variances is a good indicator of quantum chaos. As shown in Appendix~\ref{sec:breakdown} for the 3D Anderson model deep in the localized regime [see Fig.~\ref{figA2}(a) and a related discussion], the variances of diagonal and off-diagonal matrix element decay $\propto 1/V$ for $\underline{\hat{m}}_{0}$, but their ratio is clearly smaller than the RMT prediction $\Sigma^2=2$.

Finally, we would like to mention that the calculation of the variance of diagonal matrix elements of the kinetic energy $\underline{\hat{T}}$ requires the removal of the linear structure, as was discussed in Sec.~\ref{sec:structure}. Although not shown, we have confirmed that after the moving average $\overline{\underline{T}_{\alpha\alpha}}$ is removed, the square of the variance of the diagonal matrix elements $\underline{T}_{\alpha\alpha}-\overline{\underline{T}_{\alpha\alpha}}$ is $\propto 1/V$. Furthermore, the ratio of variances $\Sigma^2\approx 2$ for $V>20$, see also Fig.~\ref{fig9}.

\section{Distributions of matrix elements} \label{sec:distributions}

In this section, we study the distributions of matrix elements of observables. In Sec.~\ref{sec:nongaussian}, we 
show that the probability density functions (PDFs) of the matrix elements of $\underline{\hat n}$, $\underline{\hat h}$, and $\underline{\hat m}_0$ are non-Gaussian. Specifically, we find them to be described by the chi-square distribution with degree one, the Bessel function of the second kind, and the exponential distribution. In Sec.~\ref{sec:gaussian}, we focus on $\underline{\hat T}$, which exhibits a Gaussian distribution. We also discuss the conditions that are necessary for an observable to exhibit a Gaussian distribution of matrix elements.

\subsection{Non-Gaussian distributions} \label{sec:nongaussian}

%%%%%%%%%%%%%%%%%%%%%%%%%%%%%%%% FIGURE 7
\begin{figure*}[!t]
\centering
\includegraphics[width=0.98\textwidth]{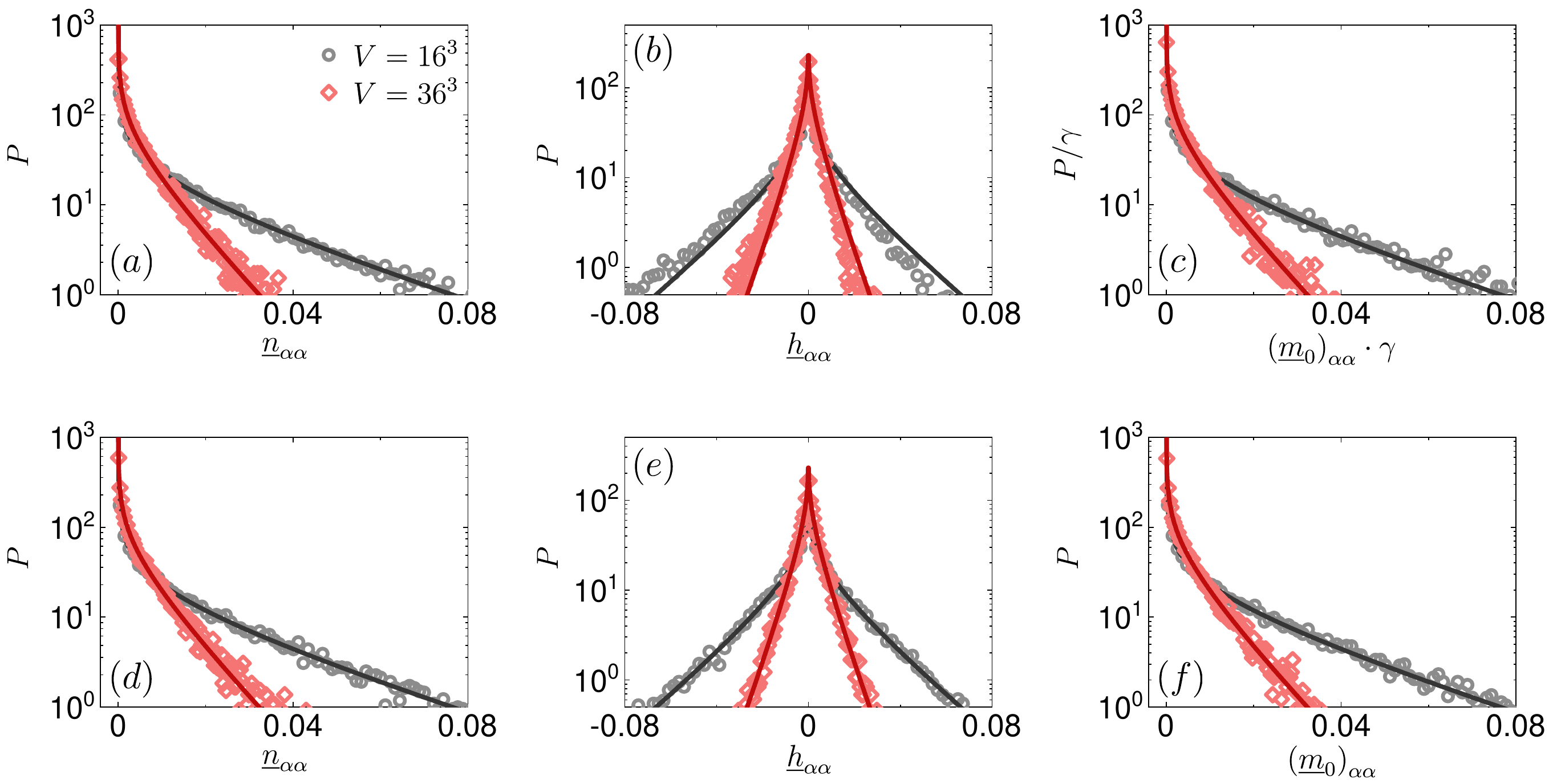}
\caption{Distributions of the diagonal matrix elements of observables in (a)--(c) the 3D Anderson model and (d)--(f) the Dirac SYK2 model. Points are numerical results for $200$ eigenstates around the mean energy, averaged over $100$ Hamiltonian realizations. The observables are: (a),(d) $\underline{\hat{n}}$, (b),(e) $\underline{\hat{h}}$, and (c),(f) $\underline{\hat{m}}_0$. Solid lines are the PDFs: (a),(d) $P_{\underline{n}_{\alpha\alpha}}$ from Eq.~(\ref{def_pdf_naa}), (b),(e) $P_{\underline{h}_{\alpha\alpha}}$ from Eq.~(\ref{def_pdf_haa}), and (c),(f) $P_{(\underline{m}_0)_{\alpha\alpha}}$ from Eq.~(\ref{def_pdf_maa}). The results in (a), (c), (d), and (f) are shifted in the $x$-axis by $1/\sqrt{V}$ (so that all the plots start at zero), while the axes in (c) are scaled by the parameter $\gamma$, which equals $442.5$ and $428.1$ for $V=16^3$ and $V=36^3$, respectively (see Sec.~\ref{sec:eigtoeig} for details).
} \label{fig7}
\end{figure*}

\begin{figure*}[!t]
\centering
\includegraphics[width=0.98\textwidth]{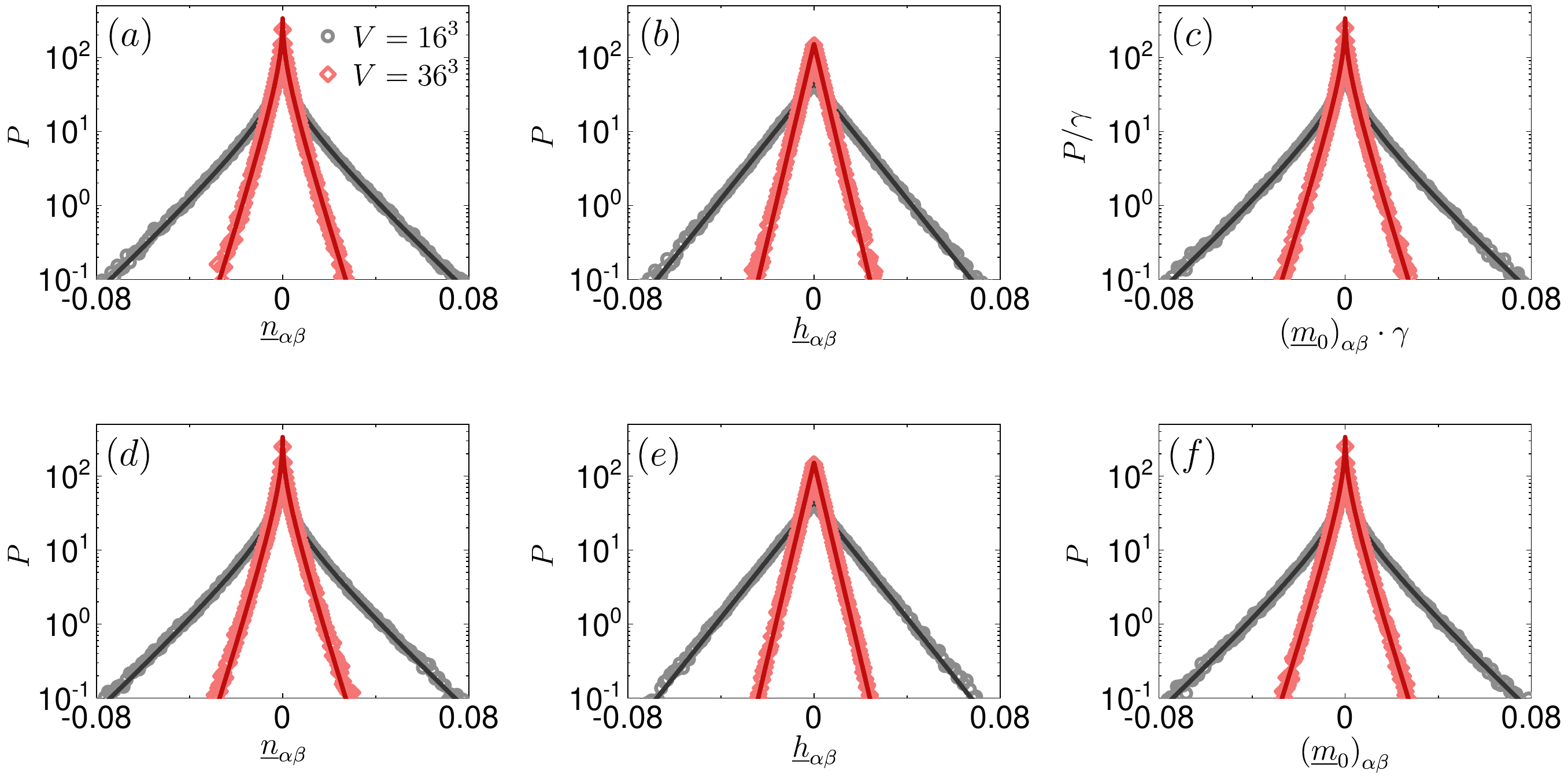}
\caption{Distributions of the off-diagonal matrix elements of observables in (a)--(c) the 3D Anderson model and (d)--(f) the Dirac SYK2 model. Points are numerical results for $200$ eigenstates around the mean energy, averaged over $20$ Hamiltonian realizations. The observables are: (a),(d) $\underline{\hat{n}}$, (b),(e) $\underline{\hat{h}}$, and (c),(f) $\underline{\hat{m}}_0$. Solid lines are the PDFs: (a),(d) $P_{\underline{n}_{\alpha\beta}}$ from Eq.~(\ref{def_pdf_nab}), (b),(e) $P_{\underline{h}_{\alpha\beta}}$ from Eq.~(\ref{def_pdf_hab}), and (c),(f) $P_{(\underline{m}_0)_{\alpha\beta}}$ from Eq.~(\ref{def_pdf_mab}). The axes in (c) are scaled by the parameter $\gamma$, which equals $441.3$ and $425.9$ for $V=16^3$ and $V=36^3$, respectively (see Sec.~\ref{sec:eigtoeig} for details).
}\label{fig8}
\end{figure*}

Here we focus on the site occupation $\underline{\hat n}$, the next-nearest neighbor correlation $\underline{\hat h}$, and the quasi-momentum occupation $\underline{\hat m}_0$, as defined in Eqs.~(\ref{def_ni})--(\ref{def_m0}). Numerical results for the distributions of the diagonal matrix elements are shown as symbols in Fig.~\ref{fig7} for the 3D Anderson model (upper row) and the Dirac SYK2 model (lower row). The corresponding numerical results for the off-diagonal matrix elements are presented in Fig.~\ref{fig8}. We consider two system sizes $V=16^3$ and $36^3$, and compute the matrix elements in 200 eigenstates around the mean energy.

The distributions of diagonal and off-diagonal matrix elements in Figs.~\ref{fig7} and~\ref{fig8} exhibit two important features. First, they are all non-Gaussian distributions. This is in striking contrast to quantum-chaotic interacting systems, for which previous studies reported solely Gaussian distributions~\cite{beugeling_moessner_15, luitz_barlev_16, khaymovich_haque_19, leblond_mallayya_19, brenes_leblond_20, brenes_goold_20, leblond_rigol_20, santos_perezbernal_20, noh_21, brenes_pappalardi_21}. Second, the numerical results in Figs.~\ref{fig7} and~\ref{fig8} (symbols) are well described by the closed-form analytical expressions (lines), which we discuss below. Recently, the analysis of the matrix elements of operators dubbed behemoths (which are nonlocal operators in many-body systems with diverging Hilbert-Schmidt norm) showed non-Gaussian distributions~\cite{khaymovich_haque_19}. The analytical forms of the distributions for the behemoths share several similarities with the distributions for one-body observables in quantum-chaotic quadratic Hamiltonians studied here.

As a first step to obtain analytical expressions for the distributions, we note that $\hat n_i$ and $\hat h_{ij}$ (reintroducing site indices for generality) have a simple structure in the single-particle site occupation basis $\{|i\rangle\}$, $\hat n_i=|i\rangle\langle i|$ and $\hat h_{ij}=|i\rangle\langle j|+|j\rangle\langle i|$, while quasi-momentum occupations $m_{\bf k}$ (introducing quasi-momentum indices for generality) have a simple structure in the single-particle quasi-momentum occupation basis $\{|{\bf k}\rangle\}$, $m_{\bf k}=|{\bf k}\rangle\langle \bf{k}|$. For the sake of keeping the discussion general for those three observables, and for others with a similar structure, let us think of our observables of interest as having a simple structure in some single-particle occupation basis $\{|\eta\rangle\}$. We then write the energy eigenstates $|\alpha\rangle$ in that basis
\begin{equation} \label{def_alpha}
|\alpha\rangle = \sum_\eta u_{\eta \alpha} |\eta\rangle \,,
\end{equation}
where $u_{\eta \alpha} = \langle \eta | \alpha \rangle$.

%%%%%%%%%%%%%%%%%%%%%%%%%%%%%%%% FIGURE 8

The key insight in the analytical derivation of the distributions is that $u_{\eta \alpha}$ behaves as a random variable drawn from a normal distribution with zero mean and variance $\sigma^2 = 1/V$. We refer to this assumption as the RMT assumption further on. As mentioned before, this allows one to show that the ratio of variances, cf.~Eq.~(\ref{def_ratio}), is $2$~\cite{dalessio_kafri_16}. Moreover, the distributions of matrix elements expressed through the probability density functions (PDFs) can be derived using the algebra for random variables. The details of the analytical calculations can be found in Appendix~\ref{sec:dist_derivations}. The main results are summarized below.

The distribution of diagonal matrix elements of the site occupation $\underline{\hat n}$ from Eq.~(\ref{def_ni}) is related to that of the square of normal random variables (see Appendix~\ref{sec:dist_derivations_p1}). It is described by a chi-square distribution with degree 1,
\begin{equation} \label{def_pdf_naa}
P_{\underline{n}_{\alpha\alpha}} (x) =\frac{V^{1/4}}{\sqrt{2\pi}} \frac{1}{\sqrt{x+\frac{1}{\sqrt{V}}}} e^{-\frac{\sqrt{V}}{2}\left[x+\frac{1}{\sqrt{V}}\right]} \,.
\end{equation}
The distribution of the corresponding off-diagonal matrix elements is related to that of the product distribution of normal random variables (see Appendix~\ref{sec:dist_derivations_p2}). It is described by a modified Bessel function of the second kind,
\begin{equation} \label{def_pdf_nab}
P_{\underline{n}_{\alpha\beta}} (x) = \frac{\sqrt{V}}{\pi}\text{K}_{0}\left(\sqrt{V}|x|\right) \,.
\end{equation}

The distribution of diagonal matrix elements of the next-nearest neighbor correlation $\underline{\hat h}$ from Eq.~(\ref{def_hij}) also follows from the product distribution of normal random variables, however with a different prefactor than the one in Eq.~(\ref{def_pdf_nab}). It reads
\begin{equation} \label{def_pdf_haa}
P_{\underline{h}_{\alpha\alpha}} (x) = \frac{1}{\pi}\sqrt{\frac{V}{2}}K_0\left(\sqrt{\frac{V}{2}}|x|\right) \,.
\end{equation}
To obtain the distribution of the corresponding off-diagonal matrix elements, one needs to calculate a sum distribution (see Appendix~\ref{sec:dist_derivations_p3}), which yields the exponential distribution
\begin{equation} \label{def_pdf_hab}
P_{\underline{h}_{\alpha\beta}} (x) = \sqrt{\frac{V}{2}} e^{-\sqrt{2V}|x|} \,.
\end{equation}

Within this framework, the distributions of the matrix elements of the quasi-momentum occupation $\underline{\hat m}_0$, Eq.~(\ref{def_m0}), are identical to the distributions of matrix elements of the site occupation in Eqs.~(\ref{def_pdf_naa}) and~(\ref{def_pdf_nab}), respectively,
\begin{align} \label{def_pdf_maa}
    P_{(\underline{m}_0)_{\alpha\alpha}} (x) & = P_{\underline{n}_{\alpha\alpha}} (x) \,, \\
    P_{(\underline{m}_0)_{\alpha\beta}} (x) & = P_{\underline{n}_{\alpha\beta}} (x) \,. \label{def_pdf_mab}
\end{align}
This is because, within the RMT assumption, both the site occupation eigenkets and the quasi-momentum occupation eigenkets are random vectors in the eigenbasis of energy eigenstates.

The agreement between our numerical results for the SYK2 model and the analytic expressions, see the lower panels in Figs.~\ref{fig7} and Fig.~\ref{fig8}, validates the correctness of our analysis and its relevance for the system sizes and averages over realizations considered. The agreement between our numerical results for the 3D Anderson model and the analytic expressions, see upper panels in Figs.~\ref{fig7} and Fig.~\ref{fig8}, is also remarkable. Below we discuss two observations about the PDFs in the 3D Anderson model.

The first one is related to the distribution of diagonal matrix elements of the next-nearest neighbor correlation $\underline{\hat h}$, see Fig.~\ref{fig7}(b). One can see that the tails of the PDF do not entirely overlap with the analytical prediction from Eq.~(\ref{def_pdf_haa}). The agreement slightly improves with increasing system size. However, even for the largest system size under investigation, $V = 36^3$, the PDF is slightly skewed towards the negative values. Larger system sizes need to be studied for this observable.

The second observation is related to the distributions of matrix elements of the quasi-momentum occupation $\underline{\hat m}_0$, see Figs.~\ref{fig7}(c) and~\ref{fig8}(c). While the agreement between the numerical and analytical results is excellent, we note that the axes are scaled by a constant $\gamma$. The origin of such a scaling was discussed in Sec.~\ref{sec:eigtoeig}. It is a consequence of anomalously large matrix elements at and close to the ground state, which make other matrix elements smaller due to the fixed Hilbert-Schmidt norm~(\ref{def_norm}). Interestingly, even though such anomalously large matrix elements in the vicinity of the ground state may suggest that the RMT treatment is not appropriate for $\underline{\hat m}_0$, the distributions presented in Figs.~\ref{fig7}(c) and~\ref{fig8}(c) show that this is not the case.

\subsection{Gaussian distributions} \label{sec:gaussian}

The results in the previous section make clear that some of the experimentally accessible local and nonlocal observables, which are traditionally studied in many-body systems, do not exhibit Gaussian distributions of diagonal and off-diagonal matrix elements in single-particle eigenstates of quantum-chaotic quadratic Hamiltonians. Our analysis also shows why this is the case, namely, the matrix elements of these observables are ``too simple'' when written in terms of random amplitudes $u_{\eta\alpha}$, where $\eta$ marks states from a certain occupation basis, while $\alpha$ marks states from the Hamiltonian eigenbasis. This is very different from what happens in the many-body eigenstates of quantum-chaotic interacting Hamiltonians.

%%%%%%%%%%%%%%%%%%%%%%%%%%%%%%%% FIGURE 9
\begin{figure}[!t]
\centering
\includegraphics[width=0.98\columnwidth]{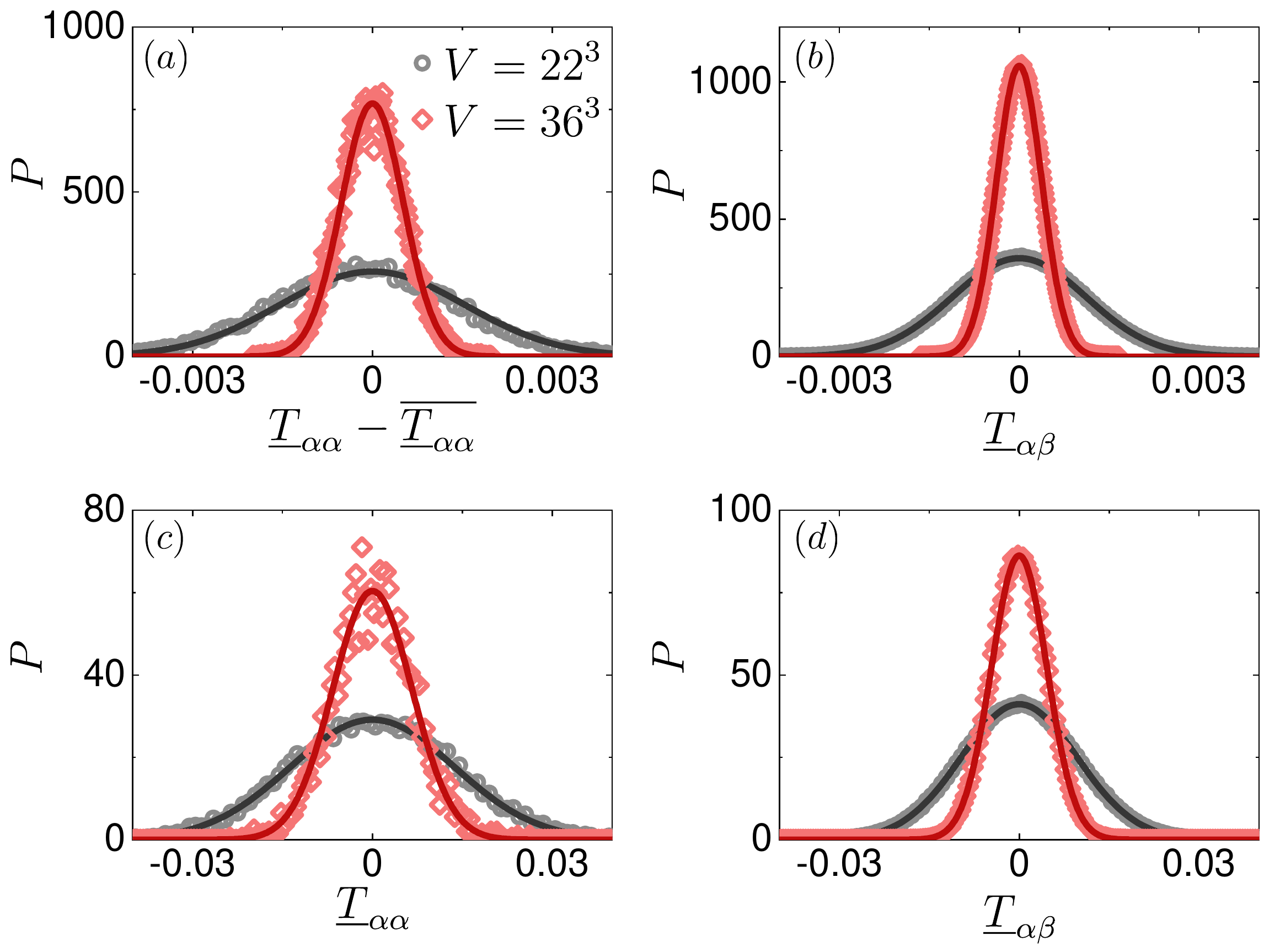}
\caption{Distributions of (a) diagonal and (b) off-diagonal matrix elements of the observable $\underline{\hat T}$ in the 3D Anderson model, and distributions of (c) diagonal and (d) off-diagonal matrix elements of the same observable in the Dirac SYK2 model. We removed the moving average from the diagonal matrix elements in the 3D Anderson model prior to the calculation of their distribution. Points are numerical results for $200$ eigenstates near the mean energy, averaged over $100$ [$20$] Hamiltonian realizations for $V=22^3$ in (a)--(d) and $V=36^3$ in (a),(b) [for $V=36^3$ in (c),(d)]. The solid lines are Gaussian distributions with zero mean and variance (a) $\sigma^2=0.013/V$, (b) $0.0066/V$, (c) $2.0/V$ and (d) $1.0/V$ for $V=36^3$. Note that the ratio of variances is $\Sigma^2\approx 2$.}
\label{fig9}
\end{figure}

To exemplify the emergence of Gaussian distributions in quantum-chaotic quadratic Hamiltonians, we study the matrix elements of the kinetic energy operator $\underline{\hat T}$ in Eq.~(\ref{def_T}). Numerical results for the distributions are shown as symbols in Fig.~\ref{fig9}, while lines are Gaussian functions with the variances computed numerically directly from the matrix elements. The agreement is excellent. It should be emphasized that we removed the linear structure from diagonal matrix elements in the 3D Anderson model prior to the calculation of their distribution.

We can understand the emergence of Gaussian distributions of the matrix elements of $\underline{\hat T}$ if we rewrite this operator in terms of its eigenvalues $-2(\cos k_x+\cos k_y+\cos k_z)$ and eigenvectors $|{\bf k}\rangle$,
\begin{equation} \label{def_Om}
    \underline{\hat T} = \frac{1}{\sqrt{6}}\sum_{\bf k} -2(\cos k_x+\cos k_y+\cos k_z)|{\bf k}\rangle\langle{\bf k}|\,.
\end{equation}
We then see that Gaussian distributions of matrix elements are a consequence of the central limit theorem, which is satisfied because the overlaps $\langle \bf{k}|\alpha\rangle$ behave as random variables and $\underline{\hat T}$ is an extensive (in $V$) sum of projector operators $|{\bf k}\rangle\langle{\bf k}|$.

More generally, we expect Gaussian distributions to emerge in observables of the form:
\begin{equation}
\label{def_g}
    \underline{\hat{g}}=\frac{1}{\sqrt{\mathcal{N}}}
    \left(\sum_{i,j=1}^{V} \kappa_{ij}\hat{c}_{i}^{\dagger}\hat{c}_{j}-\mathcal{T}\right)\,,
\end{equation}
where $\kappa_{ij}=\kappa_{ji}$ are real numbers, $\mathcal{N}=\sum_{ij}\kappa_{ij}^2/V-\sum_{ij}\kappa_{ii}\kappa_{jj}/V^2$, and $\mathcal{T}=\sum_{i=1}^{V} \kappa_{ii}$. It is, of course, needed that the eigenvalues $\{{\underline g}_\mu\}$ of $\underline{\hat{g}}$ do not have any special structure that may render the central limit theorem inapplicable. Note that the creation $\hat c^\dagger_i$ and annihilation $\hat c_i$ operators in the site occupation basis in Eq.~(\ref{def_g}) can be replaced by creation and annihilation operators in a different basis, provided that the eigenstates in the latter basis have sufficiently random overlaps with the eigenstates of the Hamiltonian. We report results for some specific local and nonlocal realizations of $\underline{\hat{g}}$ in Appendix~\ref{app:g}.

The central limit theorem emerges naturally in many-body systems even for the one-body observables studied in Sec.~\ref{sec:nongaussian}, i.e., the observables for which the matrix elements do not exhibit Gaussian distributions in single-particle energy eigenstates. In many-body systems, many-body energy eigenkets $|\tilde\alpha\rangle$ can be written as $|\tilde\alpha\rangle = \sum_{\tilde{m}} u_{\tilde{m}\tilde{\alpha}} |\tilde{m}\rangle$, where $|\tilde{m}\rangle$ are many-body basis kets. For concreteness, let us consider the local operator $\hat{n}$ in the many-body site occupation basis, where $|\tilde{m}\rangle = \prod_{\{m_j\}} \hat c_{m_j}^\dagger |\emptyset\rangle$, and $\{m_j\}$ are the occupied sites in $|\tilde{m}\rangle$. For the diagonal matrix elements, we have
\begin{equation} \label{def_nj_manybody}
    \bra{\tilde\alpha} \hat n_i \ket{\tilde\alpha} = \sum_{\tilde{m}}   \langle \tilde{m}|\hat n_i|\tilde{m}\rangle u_{\tilde{m}\tilde{\alpha}}^2 \,.
\end{equation}
Assuming that $u_{\tilde{m}\tilde{\alpha}}$ is a normally distributed random variable, and noticing that $\langle \tilde{m}|\hat n_i|\tilde{m}\rangle$ equals 1 for an exponentially large (in $V$) number of states $|\tilde{m}\rangle$, explains why the distribution of diagonal matrix elements is Gaussian. A similar analysis can be carried out for the off-diagonal matrix elements and for other one-body observables.

\section{Summary and discussion} \label{sec:conclusions}

We studied two paradigmatic quantum-chaotic quadratic Hamiltonians, the Dirac SYK2 model and the 3D Anderson model at weak disorder. Focusing on the matrix elements of observables in single-particle eigenstates, we showed that they exhibit eigenstate thermalization. Namely, that: (i) the variance of diagonal matrix elements is proportional to the inverse single-particle Hilbert space dimension (to $1/V$), and (ii) the ratio between the variance of diagonal and off-diagonal matrix elements is $2$. On the other hand, we demonstrated that the traditionally studied one-body observables (with Gaussian distributions of matrix elements in many-body eigenstates of quantum-chaotic interacting systems) can exhibit non-Gaussian distributions of matrix elements in single-particle eigenstates of quantum-chaotic quadratic models.

While it is remarkable to observe eigenstate thermalization in single-particle eigenstates of the 3D Anderson model, it is important to emphasize some differences between the thermodynamic limit in the single-particle and many-body cases. An important difference is that the density is not fixed in the former, namely, the average site occupation in single-particle eigenstates vanishes in the thermodynamic limit. The same fate befalls ${\cal O}(\bar E)$ in Eq.~\eqref{def_eth_ansatz} for the observables $\hat{n},\,\hat{h},$ and $\hat{m}_{0}$ and their normalized versions. On the other hand, ${\cal O}(\bar E)$ does not vanish for certain observables that are extensive sums of single-particle operators, such as $\hat{T}$. Those observables behave like the ones traditionally studied in the context of ETH in many-body interacting systems. The second difference is the dimension of the Hilbert space, and everything that scales with it or has a structure related to it. In the single-particle case the dimension of the Hilbert space is the number of lattice sites, while in the many-body case it is exponential in the number of lattice sites, e.g., ${\cal D}=2^V$ for qubit based systems. This leads to a different scaling of variances with a system size in both cases. Furthermore, the equivalent of ${\cal F}(\bar E, \omega)$ in Eq.~\eqref{def_eth_ansatz} for single-particle systems lacks the frequency scales available in many-body systems, in which the level spacing is $\propto1/{\cal D}$.

\acknowledgements
We acknowledge discussions with M. Mierzejewski.
This work was supported by the the Slovenian Research Agency (ARRS), Research core fundings Grants No.~P1-0044 and No.~J1-1696 (P.\L.~and L.V.) and by the National Science Foundation, Grant No.~2012145 (Y.Z. and M.R.).

\appendix

\section{Eigenstate-to-eigenstate fluctuations in Anderson insulator} \label{sec:breakdown}

In the main text, we studied the 3D Anderson model in the quantum-chaotic regime by setting $W=1$ in Eq.~(\ref{eq_HA}). Here, we complement this analysis by studying the 3D Anderson model in the localized regime. The Anderson localization transition takes place at $W_{\rm c} \approx 16.5$~\cite{slevin_ohtsuki_18}. We set the disorder strength to $W=35$, so that the main effects of localization are robust already in finite systems~\cite{kramer_mackinnon_93, markos_06, suntajs_prosen_21}.

We first study the eigenstate-to-eigenstate fluctuations of diagonal matrix elements of normalized observables, which were introduced in Sec.~\ref{sec:eigtoeig}. In the quantum-chaotic regime at $W=1$, the average $\langle\langle \delta \underline{O}_{\rm av} \rangle\rangle$ decays as $\propto 1/V^{1/2}$, see Fig.~\ref{fig5}(a), and the maximal value $\langle\langle \delta \underline{O}_{\rm max} \rangle\rangle$ decays as $\propto 1/V^\zeta$ with $0<\zeta<0.5$, see Fig.~\ref{fig5}(b).

%%%%%%%%%%%%%%%%%%%%%%%%%%%%%%%% FIGURE A1
\begin{figure}[!t]
\centering
\includegraphics[width=0.98\columnwidth]{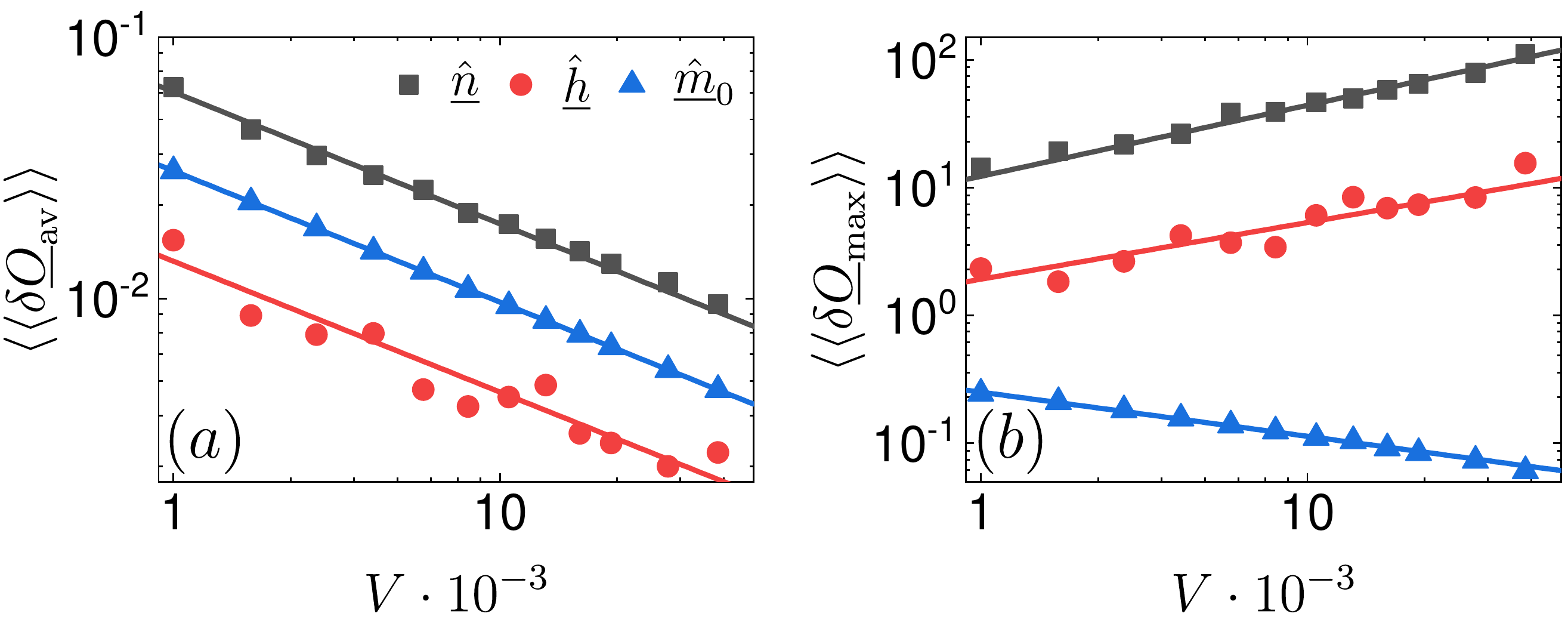}
\caption{(a) Average and (b) maximal eigenstate-to-eigenstate fluctuations in the 3D Anderson model at $W=35$. Systems with $V<36^3$ ($V=36^3$) have been averaged over $100$ ($20$) Hamiltonian realizations. Lines are $a/V^\zeta$ fits to the numerical results. For $\left<\left<\delta \underline{O}_\text{av}\right>\right>$, we get $\zeta=0.50$ when $\underline{\hat{O}}=\underline{\hat{n}}$ and $\underline{\hat{m}}_{0}$. Due to pronounced fluctuations, we fix $\zeta=0.5$ when $\underline{\hat{O}}=\underline{\hat{h}}$. For $\left<\left<\delta \underline{O}_\text{max}\right>\right>$, we get $\zeta=-0.56$, $-0.44$ and $0.35$ when $\underline{\hat{O}}=\underline{\hat{n}}$, $\underline{\hat{h}}$ and $\underline{\hat{m}}_{0}$, respectively.}\label{figA1}
\end{figure}

In Fig.~\ref{figA1}, we plot the eigenstate-to-eigenstate fluctuations of the observables $\underline{\hat n}$, $\underline{\hat h}$ and $\underline{\hat m}_0$ at $W=35$. The average fluctuations $\langle\langle \delta \underline{O}_{\rm av} \rangle\rangle$ decay for all three observables $\propto 1/V^\zeta$, see Fig.~\ref{figA1}(a), where $\zeta=0.50$ for $\underline{\hat n}$ and $\underline{\hat m}_0$, and $\zeta \approx 0.5$ for $\underline{\hat h}$ (see the figure caption for details). However, the system-size dependence of the maximal fluctuations $\langle\langle \delta \underline{O}_{\rm max} \rangle\rangle$ is drastically different for $\underline{\hat n}$ and $\underline{\hat h}$, see Fig.~\ref{figA1}(b), where $\zeta < 0$. This is a clear signature of the breakdown of the ETH in the localized regime.

The numerical results presented in Fig.~\ref{figA1} highlight the need to study both $\langle\langle \delta \underline{O}_{\rm av} \rangle\rangle$ and $\langle\langle \delta \underline{O}_{\rm max} \rangle\rangle$ in ETH analyses. This can be illustrated using as example the site occupation operator $\underline{\hat n} \equiv \underline{\hat n}_i$. Deep in the Anderson insulating regime, one can consider two classes of diagonal matrix elements:
(i) those that are nonvanishing, which correspond to localized eigenstates that are peaked at (or very close to) the site $i$;
(ii) those that are vanishing, for which the site index $i$ belongs to the tails of the localized eigenstates.
The latter class represents the overwhelming majority of matrix elements, and hence it governs the behavior of the average fluctuations $\langle\langle \delta \underline{O}_{\rm av} \rangle\rangle$. Consequently, $\langle\langle \delta \underline{O}_{\rm av} \rangle\rangle$ are small and decay with increasing $V$. On the other hand, the maximal fluctuations $\langle\langle \delta \underline{O}_{\rm max} \rangle\rangle$ are governed by the difference between the occupation at the peaks and the tails of the localized orbitals, which diverge $\propto V^{1/2}$ for the normalized operator $\underline{\hat n}$ defined in Eq.~(\ref{def_ni}), as observed in Fig.~\ref{figA1}(b).

%%%%%%%%%%%%%%%%%%%%%%%%%%%%%%%% FIGURE A2
\begin{figure}[!t]
\centering
\includegraphics[width=0.98\columnwidth]{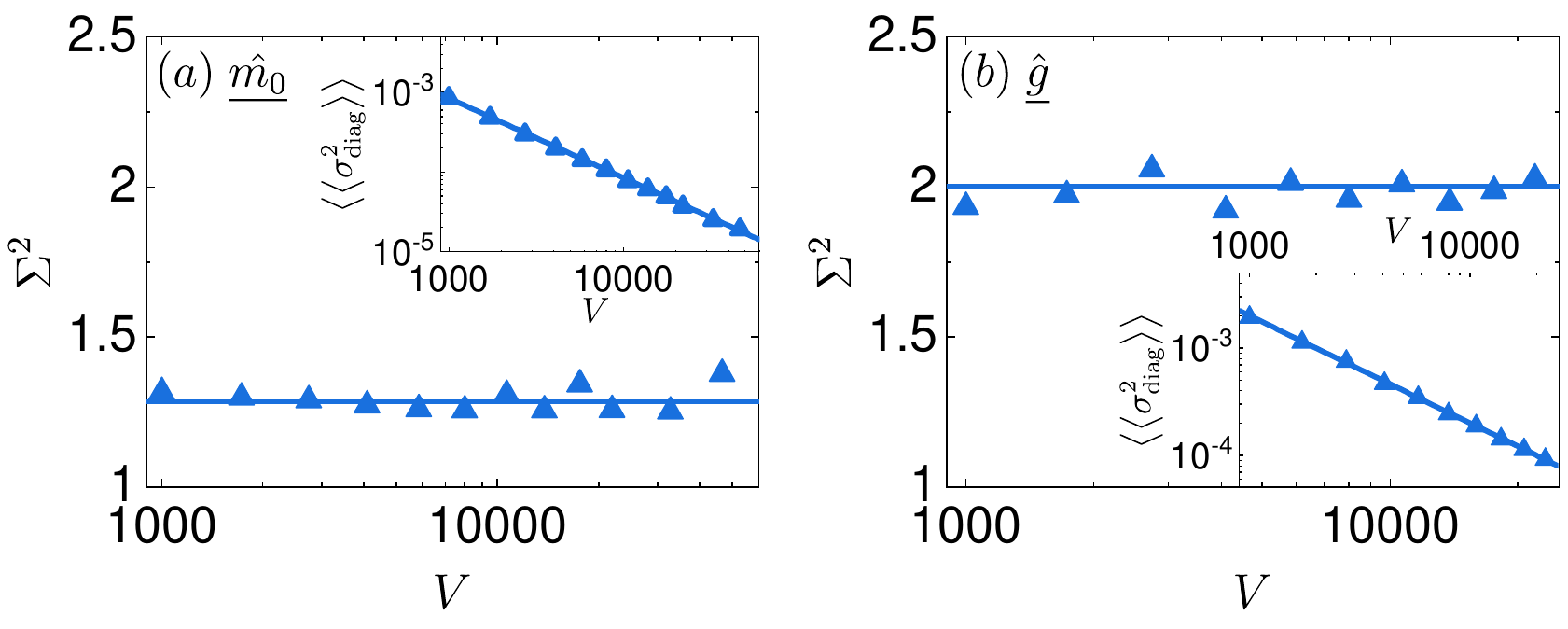}
\caption{The ratio of variances $\Sigma^2$ of the observables (a) $\underline{\hat{m_0}}$ and (b) $\underline{\hat{g}}$ in the 3D Anderson model. The all-to-all couplings $\kappa_{ij}$ in $\underline{\hat{g}}$ are normally distributed numbers with zero mean and variance $\sigma^2=2/V$ ($\sigma^2=1/V$) for diagonal (off-diagonal) matrix elements. Points are numerical results for $200$ eigenstates near the mean energy. The averaging is carried out over $100$ [$20$] Hamiltonian realizations for $V<36^3$ in (a) [$V=36^3$ in (a) and all system sizes in (b)]. The insets show the variances of the diagonal matrix elements as functions of the system size. The line in (a) is a $a/V$ fit to the numerical results with $a=0.85$, while the line in (b) is $2/V$.}
\label{figA2}
\end{figure}

Identifying the breakdown of the RMT description of the  matrix elements of the zero quasi-momentum occupation $\underline{\hat m}_0$ in the Anderson insulator requires a more detailed analysis. Figure~\ref{figA1}(b) shows that for this observable the maximal fluctuations $\langle\langle \delta \underline{O}_{\rm max} \rangle\rangle$ decay with increasing the system size, similar to the maximal fluctuations at $W=1$, see Fig.~\ref{fig5}. Moreover, the variance of diagonal matrix elements defined in Eq.~(\ref{def_variance_diag}) decays as $\langle\langle \sigma_{\rm diag}^2\rangle\rangle \propto 1/V$, as shown in the inset of Fig.~\ref{figA2}(a). These results are not surprising because the eigenkets of $\underline{\hat m}_0$ are delocalized in the eigenbasis of $\hat H$. The breakdown of the RMT description is apparent when one computes the ratio of variances in Eq.~(\ref{def_ratio}), which is shown in the main panel of Fig.~\ref{figA2}(a). One can see there that it does not approach the value $\Sigma^2=2$ predicted by the RMT.

One can, of course, always construct single-particle nonlocal operators for which the RMT description is valid even if the Hamiltonian is not quantum chaotic. To show this, we compute the matrix elements of the operator $\underline{\hat g}$ from Eq.~(\ref{def_g}) in the Anderson localized regime. The all-to-all couplings $\kappa_{ij}$ in $\underline{\hat g}$ are normally distributed random numbers, such that the operator can be seen as an independent realization of the SYK2 Hamiltonian~(\ref{eq_Hsyk2}), but traceless and properly normalized (the same operator is also studied in Appendix~\ref{sec:nonlocal}). In Fig.~\ref{figA2}(b) we show that the matrix elements of $\underline{\hat g}$ comply with the ETH: the variances of diagonal matrix elements decay as $\langle\langle \sigma^2_{\rm diag} \rangle\rangle \propto 1/V$, and the ratio of variances equals $\Sigma^2 \approx 2$. The latter is very close to $2$ already in relatively small systems with $V = 10^3$ lattice sites. This is a direct consequence of the fact that the projections of eigenstates of $\underline{\hat{g}}$ onto energy eigenstates are random numbers [see Eq. (\ref{def_g}) and recall that energy eigenstates correspond to site occupation eigenstates in the limit of infinite disorder].

\section{Off-diagonal matrix elements} \label{sec:offdiag_appendix}

\begin{figure}[!h]
\centering
\includegraphics[width=0.98\columnwidth]{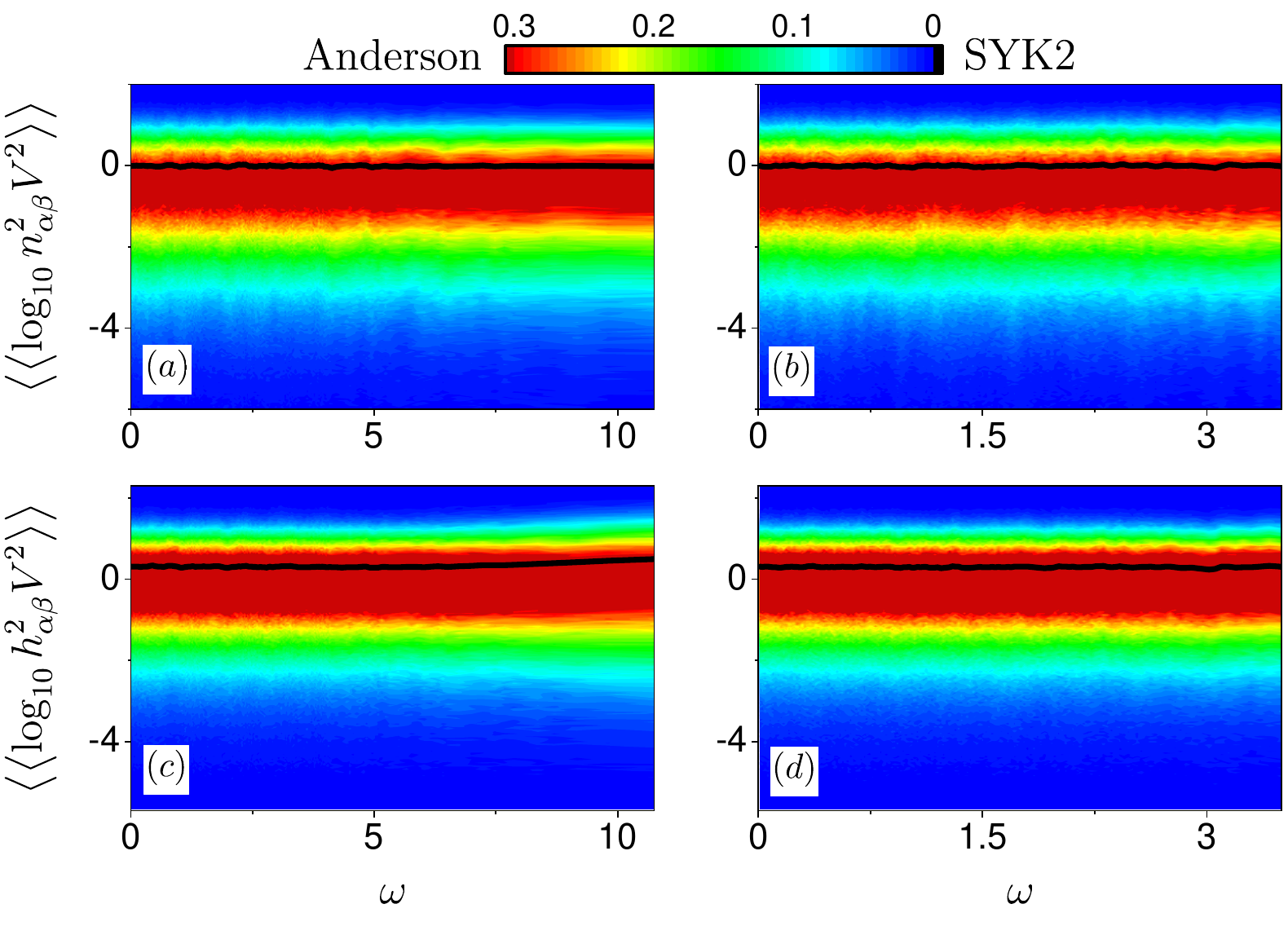}
\caption{Density plots of the off-diagonal matrix elements of the observables (a),(b) $\hat{n}$ and (c),(d) $\hat{h}$ as functions of the energy difference $\omega=|E_\alpha-E_\beta|$.  We consider $V=20^3$. The black lines denote the moving averages $\langle\langle \log_{10} \overline{ n_{\alpha\beta}^2} V^2\rangle\rangle$ and $\langle\langle \log_{10} \overline{ h_{\alpha\beta}^2} V\rangle\rangle$. Results for the 3D Anderson  model  (left  column)  and  the  Dirac  SYK2  model  (right column) have been established in the same protocol as results in Fig.~\ref{fig3}. }
\label{figA3}
\end{figure}

In Sec.~\ref{sec:structure_offdiag} we showed results for the off-diagonal matrix elements of the observables $\hat m_0$ and $\hat T$. Here we complement those results with density plots of the off-diagonal matrix elements of observables $\hat n$ and $\hat h$, shown in Fig.~\ref{figA3}. The squares of matrix elements are multiplied by $V^2$ to ensure that the corresponding moving averages (black solid lines in Fig.~\ref{figA3}) do not change when changing $V$. For both observables, one can see that there is almost no dependence on $\omega$, and the results in the 3D Anderson model (left column in Fig.~\ref{figA3}) are very similar to the results in the Dirac SYK2 model (right column in Fig.~\ref{figA3}).

%%%%%%%%%%%%%%%%%%%%%%%%%%%%%%%% FIGURE A3

\section{Variances over Hamiltonian realizations}
\label{sec:realizations}

%%%%%%%%%%%%%%%%%%%%%%%%%%%%%%%% FIGURE A7
\begin{figure}[!t]
\centering
\includegraphics[width=\columnwidth]{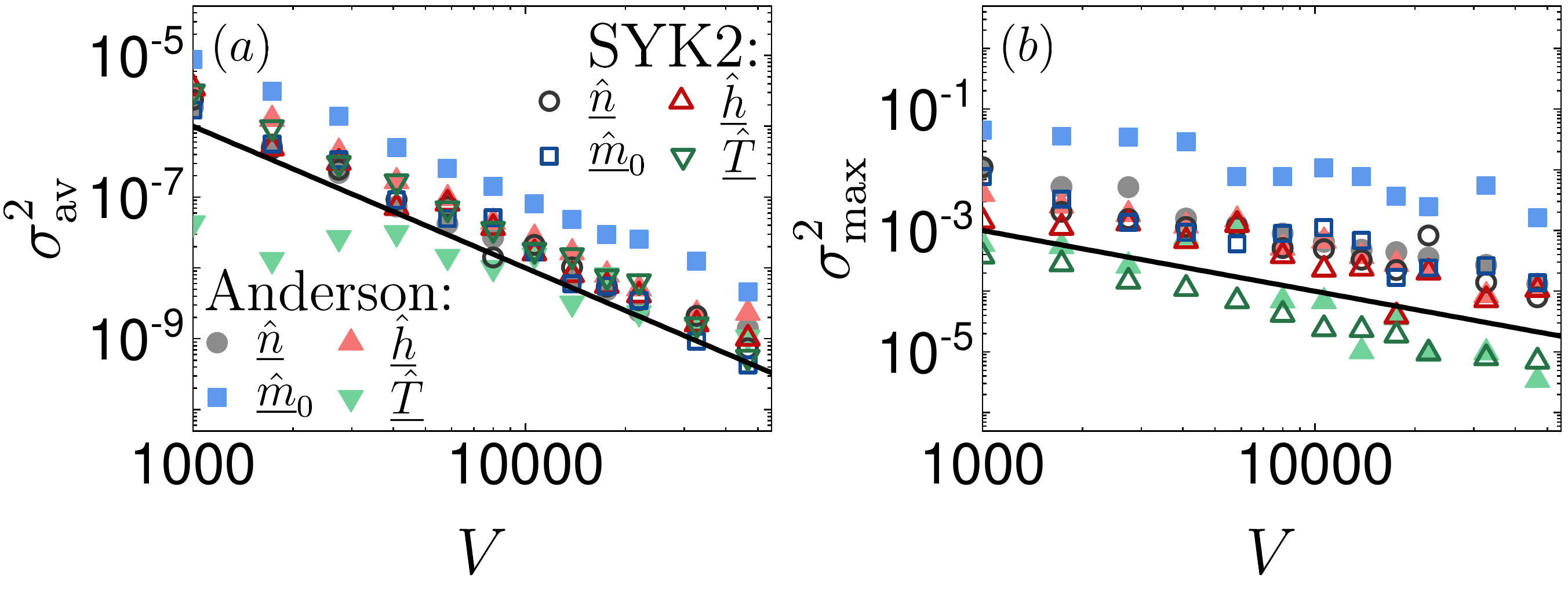}
\caption{Variances of (a) the average and (b) the maximal eigenstate-to-eigenstate fluctuations over Hamiltonian realizations. Filled (open) symbols correspond to the numerical results for the 3D Anderson (Dirac SYK2) model. $\delta O_\text{av}$ and $\delta O_\text{max}$ are calculated from $80\%$ ($100$) of the single-particle eigenstates in the middle of the spectrum for $\underline{\hat{n}}$, $\underline{\hat{h}}$, and $\underline{\hat{m}_0}$ ($\underline{\hat{T}}$). $100$ ($20$) Hamiltonian realizations are considered for $V<28^3$ ($V\ge 28^3$). In the 3D Anderson model, $\sigma^2_\text{av}$ and $\sigma^2_\text{max}$ of $\underline{\hat{m}}_0$ are multiplied by $\gamma^2$, while $\sigma^2_\text{av}$ of $\underline{\hat{T}}$ is multiplied by $64$. The latter is consistent with Fig.~\ref{fig5}. For clarity, no multiplication is done for $\sigma^2_\text{max}$ of $\underline{\hat{T}}$. Lines in (a) and (b) correspond to $1/V^2$ and $1/V$, respectively.}
\label{figA7}
\end{figure}

In the main text, we considered indicators of eigenstate thermalization in the single-particle sector of the 3D Anderson and Dirac SYK2 models. The results were averaged over different Hamiltonian realizations.
Here we study the variances over Hamiltonian realizations for the eigenstate-to-eigenstate fluctuations studied in Sec.~\ref{sec:eigtoeig}.
Specifically, we define the variance of the average
\begin{equation}
    \sigma_\text{av}^2=\left<\left<\delta O_\text{av}^2\right>\right>-\left<\left<\delta O_\text{av}\right>\right>^2
\end{equation}
and the variance of the maximal eigenstate-to-eigenstate fluctuations
\begin{equation}
    \sigma_\text{max}^2=\left<\left<\delta O_\text{max}^2\right>\right>-\left<\left<\delta O_\text{max}\right>\right>^2\;,
\end{equation}
with respect to the Hamiltonian realizations. As in Fig.~\ref{fig5} for the 3D Anderson model, we multiply $\sigma^2_\text{av}$ and $\sigma^2_\text{max}$ of $\underline{\hat{m}}_0$ by $\gamma^2$, and $\sigma^2_\text{av}$ of $\underline{\hat{T}}$ by $64$. No multiplication is done for $\sigma^2_\text{max}$ of $\underline{\hat{T}}$.

Figure~\ref{figA7} shows the variances $\sigma_{\rm av}^2$ and $\sigma_{\rm max}^2$ for observables $\underline{\hat{n}}$, $\underline{\hat{h}}$, $\underline{\hat{m}_0}$, and $\underline{\hat{T}}$. Results for the 3D Anderson model are shown using filled symbols while results for the SYK2 model are shown using open symbols. One can see that $\sigma_{\rm av}^2$ and $\sigma_{\rm max}^2$ are very small for all observables, and they decrease with increasing the system size. The decrease appears to be consistent with $\sigma_{\rm av}^2 \propto 1/V^2$ and $\sigma_{\rm max}^2 \propto 1/V$, see the lines in Fig.~\ref{figA7}. The vanishing of $\sigma_{\rm av}^2$ and $\sigma_{\rm max}^2$ in the thermodynamic limit suggest that the results reported in the main text are valid for a typical realization of the Hamiltonians under investigation.

\section{Derivation of distributions of\\ matrix elements} \label{sec:dist_derivations}

Next, we derive the closed-form expressions for the PDFs of the matrix elements of the observables $\underline{\hat n}$ and $\underline{\hat h}$ reported in Eqs.~(\ref{def_pdf_naa})--(\ref{def_pdf_hab}) of the main text. For both observables one can interpret Eq.~(\ref{def_alpha}) as the expansion of the Hamiltonian eigenstate $|\alpha\rangle$ in the site occupation basis, where $|\eta \rangle \equiv |i\rangle = \hat c^\dagger_i |\emptyset \rangle$. Note that the site occupation basis is the eigenbasis for $\underline{\hat n}_i$ but not for $\underline{\hat h}_{ij}$. Still, for the derivation of the specific distributions it is convenient to use this basis for both observables. The diagonal matrix elements of $\underline{\hat n} \equiv \underline{\hat n}_i$ are
\begin{align} \label{coef_naa}
    \underline{n}_{\alpha\alpha}
    &= \frac{V}{\sqrt{V-1}} \bra{\alpha} \hat c_i^\dagger \hat c_i^{} \ket{\alpha} - \frac{1}{\sqrt{V-1}} \nonumber \\
    &= \frac{V}{\sqrt{V-1}} u_{i\alpha}^2 - \frac{1}{\sqrt{V-1}} \,,
\end{align}
and the off-diagonal matrix elements are
\begin{equation} \label{coef_nab}
    \underline{n}_{\alpha\beta} = \frac{V}{\sqrt{V-1}} u_{i\alpha} u_{i\beta} \,.
\end{equation}
To simplify the expressions in Eqs.~(\ref{coef_naa}) and~(\ref{coef_nab}),  we replace $V-1 \to V$ having in mind $V \gg 1$. The diagonal matrix elements of $\underline{\hat h} \equiv \underline{\hat h}_{ij}$ are
\begin{align} \label{coef_haa}
    \underline{h}_{\alpha\alpha}
    &= \sqrt{\frac{V}{2}} \bra{\alpha} \hat c_i^\dagger \hat c_j^{} + c_j^\dagger \hat c_i^{} \ket{\alpha} \nonumber \\
    &= \sqrt{\frac{V}{2}} ( u_{i\alpha} u_{j \alpha} + u_{j \alpha} u_{i \alpha} )
    = \sqrt{2V} u_{i\alpha} u_{j \alpha} \,,
\end{align}
and the off-diagonal matrix elements are
\begin{align} \label{coef_hab}
    \underline{h}_{\alpha\beta}
    &= \sqrt{\frac{V}{2}} ( u_{i\alpha} u_{j \beta} + u_{j \alpha} u_{i \beta} ) \,.
\end{align}

The starting point for the derivation of the distributions is the RMT assumption about the coefficients $u_{i\alpha}$, i.e., that they are normally distributed real random variables with zero mean and variance $\sigma^2 = 1/V$,
\begin{equation} \label{rand_gauss}
P_u\left(x\right) = \frac{1}{\sqrt{2\pi\sigma^2}}\exp\left(-\frac{x^2}{2\sigma^2}\right) \,.
\end{equation}
Below we review some basic results for random variables that have a direct application in the derivation of the PDFs of the matrix elements from Eqs.~(\ref{coef_naa})-(\ref{coef_hab}).

\subsection{Functions of normal random variables} \label{sec:dist_derivations_p1}

Let $u$ be a random variable and $v = g(u)$ be a function of $u$. If $g$ is differentiable and invertible, such that $u = h(v)$ with $h = g^{-1}$, the PDF of $v$ can be written as
\begin{equation}
    P_v(y) = P_u \left( h(y) \right) \left| \frac{dh(y)}{dy} \right| \,.
\end{equation}
For example, if $v = g(u) = a u$, with $a$ being a constant, it follows that $h(v) = v/a$ and $|dh(v)/dv| = 1/|a|$, such that
\begin{equation} \label{rand_Pv_lin1}
    P_v(y) = \frac{1}{|a|} P_u \left( \frac{y}{a} \right) \,.
\end{equation}
If one further adds a constant $b$ to $v$ such that $v = g(u) = a u + b$, then $h(v) = (v-b)/a$ and
\begin{equation} \label{rand_Pv_lin2}
    P_v(y) = \frac{1}{|a|} P_u \left( \frac{y-b}{a} \right) \,.
\end{equation}
One the other hand, if $g$ is not invertible but there exist a finite number of $x_i$ such that $y = g(x_i)$ [$x_i$ and $y$ are possible outcomes of $u$ and $v$, respectively], then
\begin{equation} \label{rand_func2}
    P_v(y) = \sum_i \left| \frac{dg_i^{-1}(y)}{dy} \right| P_u(g_i^{-1}(y)) \,.
\end{equation}
A simple illustration of the latter case is the function $v = g(u) = u^2$, for which $u = \pm \sqrt{v} = g_{1,2}^{-1}(v)$ so $|dg_{1,2}^{-1}(v)/dv| = 1/(2\sqrt{v})$. Using Eq.~(\ref{rand_func2}) one gets
\begin{equation}
    P_v(y) = \frac{1}{2\sqrt{y}} P_u(\sqrt{y}) + \frac{1}{2\sqrt{y}} P_u(-\sqrt{y}) \,.
\end{equation}
If $P_u$ is a Gaussian function as in Eq.~(\ref{rand_gauss}), one gets
\begin{equation} \label{rand_Pv_chi2}
    P_v(y) = \frac{1}{\sqrt{y}} \frac{1}{\sqrt{2\pi \sigma^2}} \exp\left( -\frac{y}{2\sigma^2} \right) \,,
\end{equation}
which is also known as the chi-square distribution $\chi^2_k$ with degree $k=1$.

\underline{Application}.
The distribution $P_{\underline{n}_{\alpha\alpha}}(x)$ of the diagonal matrix elements $\underline{n}_{\alpha\alpha}$ in Eq.~(\ref{coef_naa}) is derived using first Eq.~(\ref{rand_Pv_chi2}) followed by Eq.~(\ref{rand_Pv_lin2}), and results in Eq.~(\ref{def_pdf_naa}) in the main text.

\subsection{Product distribution of normal random variables} \label{sec:dist_derivations_p2}

Let $u$ and $u'$ be two independent random variables with the corresponding PDFs $P_u(x)$ and $P_{u'}(x')$ [$x$ and $x'$ are possible outcomes of $u$ and $u'$, respectively], and let $v = u u'$ be the product of these two variables. The product distribution of the latter is denoted as $P_v(y)$ and can be obtained as
\begin{align}
    P_v(y)
    &= \int_{-\infty}^{\infty} \int_{-\infty}^{\infty} P_u (x) P_{u'}(x') \delta(x x' - y) dx dx' \nonumber \\
    &= \int_{-\infty}^{\infty} \frac{1}{|x|} P_u(x) P_{u'}(y/x) dx \,.
\end{align}
If $P_u$ and $P_{u'}$ are both normal distributions with the same variance, then the product distribution $P_v$ is
\begin{align}
    P_v(y) &= \frac{1}{2\pi \sigma^2} \int_{-\infty}^{\infty} \frac{1}{|x|} \exp\left(- \frac{x^2 + y^2/x^2}{2\sigma^2} \right) dx \nonumber \\
    &= \frac{1}{\pi\sigma^2}K_0\left(\frac{|y|}{\sigma^2}\right) \,, \label{rand_K0}
\end{align}
where $K_0$ is the modified Bessel function of the second kind.

\underline{Applications}.
The distribution $P_{\underline{n}_{\alpha\beta}}(x)$ of the off-diagonal matrix elements $\underline{n}_{\alpha\beta}$ in Eq.~(\ref{coef_nab}) is derived using first Eq.~(\ref{rand_K0}) followed by Eq.~(\ref{rand_Pv_lin1}), and results in Eq.~(\ref{def_pdf_nab}) in the main text. Similarly, the distribution $P_{\underline{h}_{\alpha\alpha}}(x)$ of the diagonal matrix elements $\underline{h}_{\alpha\alpha}$ in Eq.~(\ref{coef_haa}) is derived using identical steps, and results in Eq.~(\ref{def_pdf_haa}) in the main text.

\subsection{Sum distributions} \label{sec:dist_derivations_p3}

Let $u$ and $u'$ be two independent random variables with the corresponding PDFs $P_u(x)$ and $P_{u'}(x')$, and let $v = u + u'$ be the sum of these two variables. The sum distribution of the latter is denoted as $P_v(y)$ and can be obtained by the convolution
\begin{equation} \label{rand_conv}
    P_v(y) = \int_{-\infty}^{\infty} P_u(x) P_{u'}(y - x) dx \,.
\end{equation}
A convenient way of calculating the sum distribution is through the so-called characteristic functions, i.e., the Fourier transforms of the PDFs. Let $R_w(q)$ be the Fourier transform of $P_w(z)$, defined as $R_w(q) = \int_{-\infty}^\infty e^{iqz} P_w(z) dz$. Since the Fourier transform of $P_v(y)$ from Eq.~(\ref{rand_conv}) can be expressed as a product of two Fourier transforms, $R_v(q) = R_u(q) R_{u'}(q)$, one can calculate $P_v$ using the relation
\begin{equation}
    P_v(y) = {\rm FT}^{-1} \left[ R_u(q) R_{u'}(q) \right] \,.
\end{equation}
If the distributions $P_u$ and $P_{u'}$ are identical and given by the modified Bessel function of the second kind from Eq.~(\ref{rand_K0}), then their characteristic function is $R_u(q) = 1/\sqrt{1 + q^2 \sigma^4}$, and the sum distribution is
\begin{align}
P_v(y)
    & =\frac{1}{2\pi}\int_{-\infty}^{\infty}\exp\left(-iqy\right)\frac{1}{1+\sigma^4q^2}dq \nonumber \\
    & =\frac{1}{2\sigma^2}\exp\left(-\frac{|y|}{\sigma^2}\right) \,. \label{rand_exp}
\end{align}

\underline{Application}.
The distribution $P_{\underline{h}_{\alpha\beta}}(x)$ of the off-diagonal matrix elements $\underline{h}_{\alpha\beta}$ in Eq.~(\ref{coef_hab}) is derived using first Eq.~(\ref{rand_exp}) followed by Eq.~(\ref{rand_Pv_lin1}), and results in Eq.~(\ref{def_pdf_hab}) in the main text.

\section{Distributions of matrix elements of observables $\underline{\hat{g}}$} \label{app:g}

In Sec.~\ref{sec:gaussian} we showed that the matrix elements of the operator $\hat T$ exhibit a Gaussian distribution. Below we consider the operator $\underline{\hat g}$ from Eq.~(\ref{def_g}), which is defined in a general form using $\kappa_{ij}$ as the coupling between the sites $i$ and $j$. In particular, we study two instances of the operator $\underline{\hat g}$ that can be viewed as independent realizations of the Dirac SYK2 and the 3D Anderson Hamiltonians. We show that in both cases the distributions of matrix elements are Gaussian.

\subsection{An independent realization of the\\ SYK2 Hamiltonian} \label{sec:nonlocal}

%%%%%%%%%%%%%%%%%%%%%%%%%%%%%%%% FIGURE A4
\begin{figure}[!t]
\centering
\includegraphics[width=0.98\columnwidth]{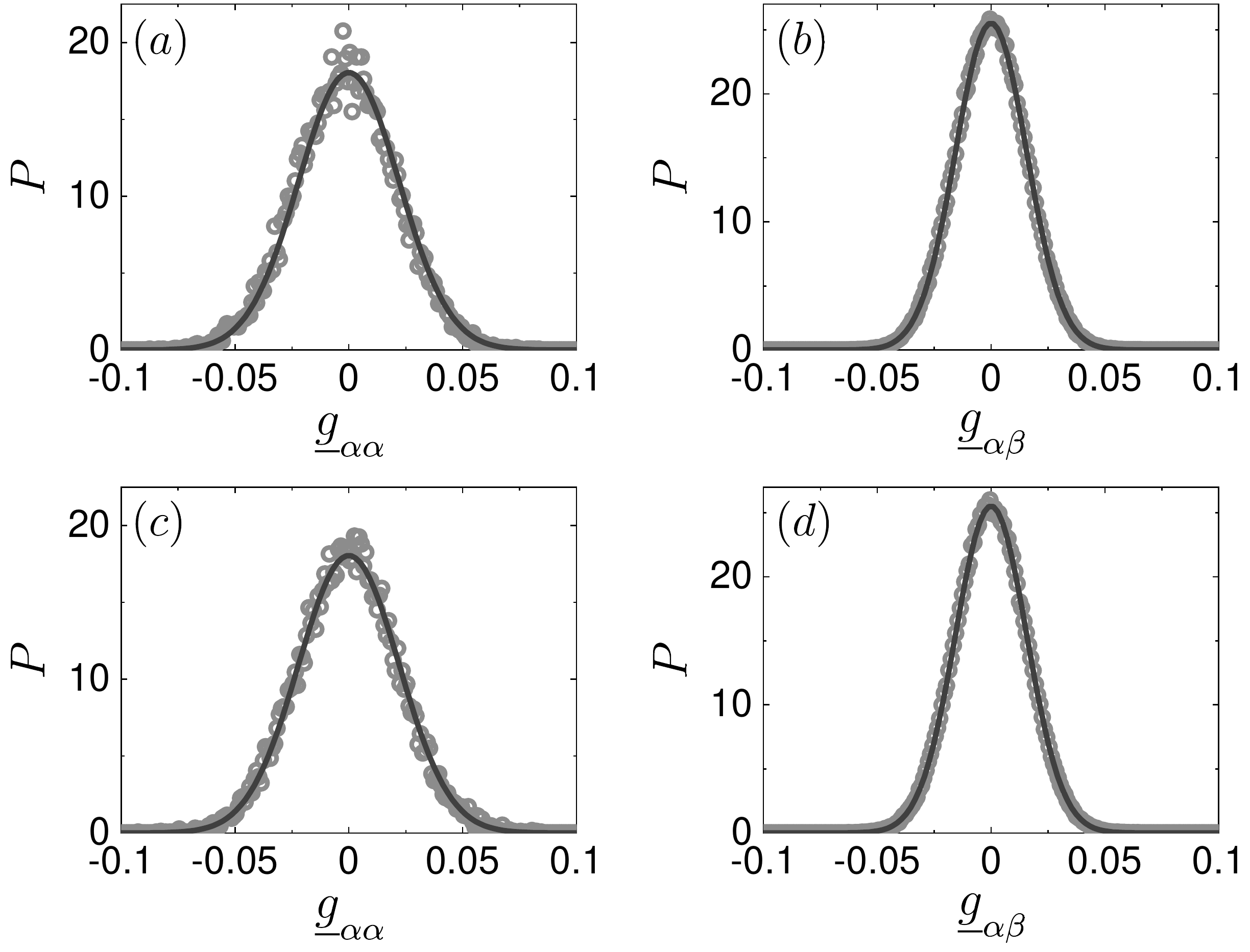}
\caption{Distributions of (a) diagonal and (b) off-diagonal matrix elements of the observable $\underline{\hat g}$ in the 3D Anderson model, and distributions of (c) diagonal and (d) off-diagonal matrix elements of the same observable in the Dirac SYK2 model. The system size is $V=16^3$. The all-to-all couplings $\kappa_{ij}$ in $\underline{\hat g}$ [see Eq.~(\ref{def_g})] are normally distributed random numbers with zero mean and variance $\sigma^2=2/V$ ($\sigma^2=1/V$) for the diagonal (off-diagonal) matrix elements. The parameters $\kappa_{ij}$ are fixed, so averages are carried out only over Hamiltonian realizations. Points are numerical results for $200$ eigenstates near the mean energy, averaged over $100$ and $20$ Hamiltonian realizations in (a),(c) and (b),(d), respectively. The solid lines are Gaussian distributions with zero mean and variance (a),(c) $\sigma^2=2/V$ and (b),(d) $\sigma^2=1/V$.}
\label{figA4}
\end{figure}

We first consider the case in which the coefficients $\kappa_{ij}$ in $\underline{\hat g}$~\eqref{def_g} are normally distributed random variables with a variance of diagonal matrix elements that is twice that of off-diagonal ones. We consider a single realization of those coefficients (defining a single observable $\underline{\hat g}$ that can be seen as an independent realization of the SYK2 Hamiltonian, but traceless and properly normalized) and carry out averages over different Hamiltonian realizations. The resulting distributions of matrix elements are shown in Figs.~\ref{figA4}(a) and~\ref{figA4}(b) for the 3D Anderson model, and in Figs.~\ref{figA4}(c) and~\ref{figA4}(d) for the Dirac SYK2 model. The PDFs are, as expected, Gaussian. We checked (not shown) that similar results are obtained if $\kappa_{ij}$ are random variables with a box distribution, or if the diagonal elements $\kappa_{ii}$ are normally distributed random variables while the off-diagonal elements are zero.

\subsection{An independent realization of the\\ 3D Anderson Hamiltonian} \label{sec:anderson_observable}

Next we consider the case in which the diagonal values $\kappa_{ii}$ in $\underline{\hat g}$~\eqref{def_g} are random variables with a box distribution in the interval $[-1/2,1/2]$, while the off-diagonal values $\kappa_{ij}$ are -1 for nearest neighbor sites in the cubic lattice and zero otherwise. This instance of the operator $\underline{\hat g}$ can be seen as an independent realization of the 3D Anderson Hamiltonian $\hat H_{\rm A}$~(\ref{eq_HA}) at $W=1$ with a unit Hilbert-Schmidt norm.

%%%%%%%%%%%%%%%%%%%%%%%%%%%%%%%% FIGURE A5
\begin{figure}[!t]
\centering
\includegraphics[width=0.85\columnwidth]{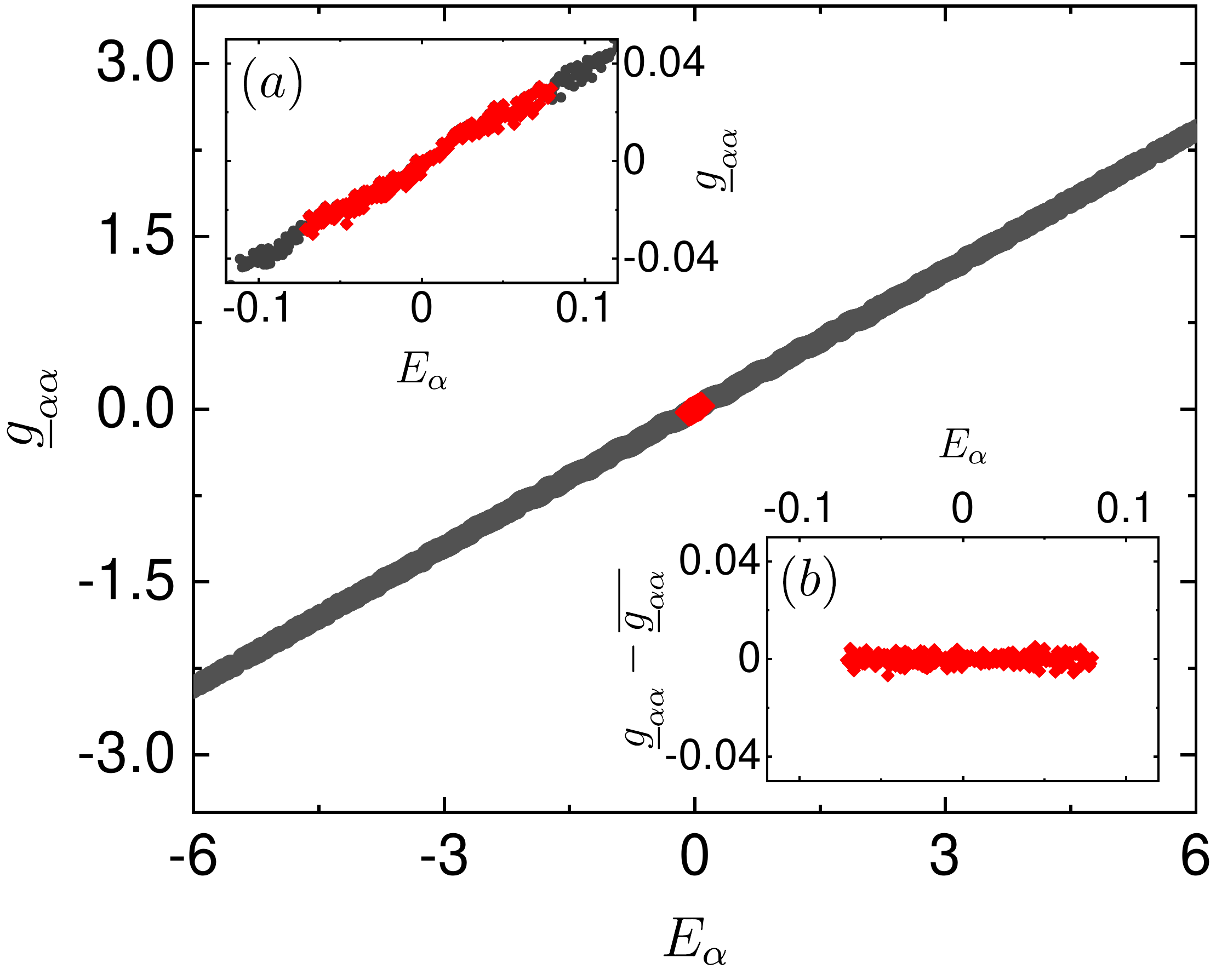}
\caption{Diagonal matrix elements of the observable $\underline{\hat g}$ in the 3D Anderson model vs the energy eigenvalue $E_\alpha$ for $V=22^3$. The diagonal $\kappa_{ii}$ in $\underline{\hat g}$ [see Eq.~(\ref{def_g})] are random numbers with a box distribution in the interval $[-1/2,1/2]$, while the off-diagonal $\kappa_{ij}$ are -1 for the nearest neighbors on the cubic lattice and zero otherwise. We study a single realization of $\underline{\hat g}$. Gray/dark (red/light) symbols are numerical results for all eigenstates ($200$ eigenstates around the mean energy). Insets: (a) A close-up of the diagonal matrix elements around the mean energy of the entire spectrum. (b) The same matrix elements as in (a) but with the moving average $\overline{\underline{g}_{\alpha\alpha}}$ subtracted. For a target eigenstate $\alpha$, the moving average $\overline{\underline{g}_{\alpha\alpha}}$ is computed using the matrix elements of the 20 closest eigenstates.}
\label{figA5}
\end{figure}

\begin{figure}[!t]
\centering
\includegraphics[width=0.98\columnwidth]{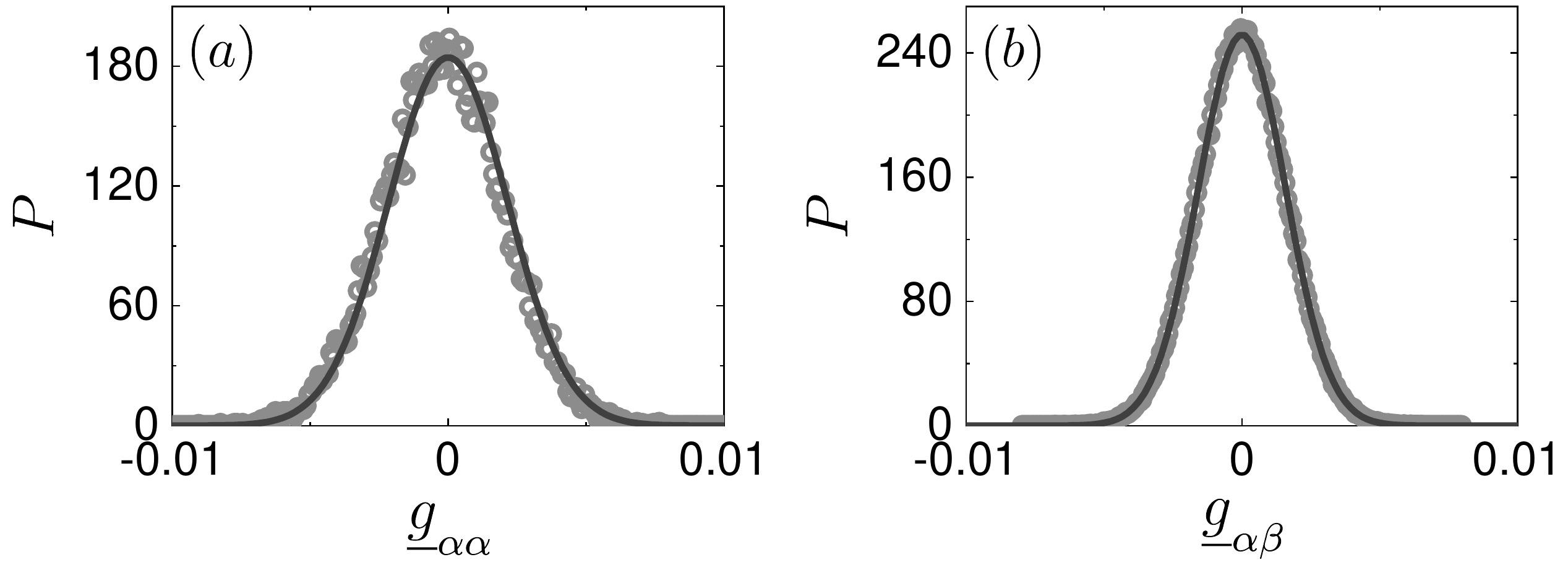}
\caption{Distributions of (a) diagonal and (b) off-diagonal matrix elements of the observable $\underline{\hat g}$ (the same observable as in Fig.~\ref{figA5}) in the 3D Anderson model for $V = 22^3$. The structure of diagonal matrix elements was removed before determining the distribution (see text, and Fig. \ref{figA5}, for details). Points are numerical results for $200$ eigenstates around the mean energy, averaged over $100$ and $20$ Hamiltonian realizations in (a) and (b), respectively. Solid lines are Gaussian distributions with zero mean and variances (a) $\sigma^2=0.0498/V$ and (b) $\sigma^2=0.0267/V$, which are calculated from the numerical results.}
\label{figA6}
\end{figure}

The main panel in Fig.~\ref{figA5} shows the diagonal matrix elements of $\underline{\hat g}$ in a single Hamiltonian realization. They can be seen to be close to, but fluctuating about, a linear function of $E_\alpha$ [similar to the diagonal matrix elements of $\hat{\underline T}$ in Fig.~\ref{fig2}(a)]. The fluctuations are more visible in Fig.~\ref{figA5}(a), where we show a close-up about the mean energy of the spectrum. When calculating the variance and distributions of diagonal matrix elements, it is important to subtract any global structure, so that the resulting matrix elements have the same mean throughout the spectrum. We achieve this in our analysis of the central $200$ eigenstates [marked with red/light color in Fig.~\ref{figA5}(a)] by subtracting the moving average, see Fig.~\ref{figA5}(b). The moving average is computed using the matrix elements for the closest $20$ states. The distribution of the structureless matrix elements is then well described by the Gaussian function, see Fig.~\ref{figA6}(a). Similar distributions were observed for the operators presented in Fig.~\ref{figA4}.

In the case of off-diagonal matrix elements, the structure in the frequency space does not play any significant role, so one can directly study the distributions obtained from $200$ eigenstates around the mean energy, as in Fig.~\ref{figA4}. The results in Fig.~\ref{figA6}(b) show a very good agreement with a Gaussian distribution. 

For both Gaussian functions in Fig.~\ref{figA6} (lines) the variance was computed numerically directly using the matrix elements. We also find, not shown, that the ratio of the variances is $\Sigma^2 \approx 1.9$ (close to the RMT result $\Sigma^2 = 2$).

%%%%%%%%%%%%%%%%%%%%%%%%%%%%%%%% FIGURE A6

\bibliographystyle{biblev1}
\bibliography{references}

\end{document}